\documentclass[%
 reprint,
 amsmath,amssymb,
 aps,
]{revtex4-2}

\usepackage{graphicx}% Include figure files
\usepackage{dcolumn}% Align table columns on decimal point
\usepackage{bm}% bold math
\usepackage[utf8]{inputenc}
\usepackage[T1]{fontenc}
\usepackage{hyperref}
\usepackage{lineno}
\usepackage{amsmath}
\usepackage{epsfig}
\usepackage{color}
\usepackage[export]{adjustbox}
\usepackage{xspace}
\usepackage[T1]{fontenc}
\newcommand{ \la }{\langle}
\newcommand{ \ra }{\rangle}

\def\PbPb  {\mbox{Pb--Pb}\xspace}

\def\snn   {\mbox{$\sqrt{s_{_{\rm NN}}}$}\xspace}
\newcommand{\Fpsi}[1]{\ensuremath{A^{\rm f}_#1}}
\newcommand{\Fvn}[1]{\ensuremath{M^{\rm f}_#1}}

\newcommand{\pt}{\ensuremath{p_{\rm T}}\xspace}
\newcommand{\pta}{\ensuremath{p_{\rm T}^{\rm a}}\xspace}
\newcommand{\ptt}{\ensuremath{p_{\rm T}^{\rm t}}\xspace}

\begin{document}

\preprint{APS/123-QED}

\title{Transverse momentum decorrelation of the flow vector in \PbPb collisions at \snn = 5.02 TeV}% Force line breaks with \\

\author{Emil Gorm Nielsen}
 \email{emil.gorm.nielsen@cern.ch}
\author{You Zhou}
 \email{you.zhou@cern.ch}
\affiliation{%
 Niels Bohr Institute, University of Copenhagen\\
}%

\date{\today}% It is always \today, today,
             %  but any date may be explicitly specified

\begin{abstract}
The individual studies of the anisotropic flow vector, flow angle and flow magnitude fluctuations with multi-particle correlations provide insight into the initial conditions and properties of the quark-gluon plasma (QGP) created in high-energy heavy-ion collisions. Recent measurements of these fluctuations have been available and the comparison to hydrodynamic models shows unique sensitivities to the initial conditions of the system, but also a puzzling dependence on the specific shear viscosity $\eta/s$. In this paper, a systematic study with A Multi-Phase Transport (AMPT) model using different tunings of the initial conditions, partonic cross section and hadronic interactions investigates the \pt-dependent flow vector, flow angle and flow magnitude fluctuations. It is found that the transport model reasonably describes the flow vector, flow angle and flow magnitude fluctuations observed in data and that the fluctuations are driven by fluctuations in the initial state. The comparison of data and model presented in this paper enables further constraints on the initial conditions of the heavy-ion collisions. 
\end{abstract}

%\keywords{Suggested keywords}%Use showkeys class option if keyword
                              %display desired
\maketitle

%\tableofcontents

\section{\label{sec:level1}Introduction}
The primary purpose of ultra-relativistic heavy-ion collisions, such as those at the Relativistic Heavy-Ion Collider (RHIC) and the Large Hadron Collider (LHC), is to create the strongly-coupled Quark-Gluon Plasma (QGP) and measure its properties \cite{BRAHMS:2004adc,STAR:2005gfr,PHENIX:2004vcz,PHOBOS:2004zne,Muller:2012zq}. The azimuthal anisotropy of produced particles is a key observable in probing the QGP \cite{Ollitrault:1992bk,Voloshin:2008dg}, and it is typically characterized by a Fourier expansion of the azimuthal angle distribution \cite{Voloshin:1994mz}
\begin{align}
P(\varphi) =\sum_{n=-\infty}^\infty\overline{V}_ne^{-in\varphi},
\end{align}
where $\varphi$ is the azimuthal angle of the produced particles and $\overline{V}_n = v_ne¨^{in\Psi_n}$ is the n-th order flow vector with magnitude $v_n$ and orientation $\Psi_n$. The magnitude $v_n$ is the n-th order flow coefficient, and the orientation $\Psi_n$ is the symmetry plane angle. The anisotropic flow has been studied extensively at RHIC \cite{STAR:2000ekf,PHENIX:2003qra,STAR:2013qio,PHENIX:2015idk} and the LHC \cite{Aamodt:2010pa,ALICE:2011ab,Abelev:2014pua,Adam:2016izf,Acharya:2017zfg,ATLAS:2012at,ATLAS:2011ah,Aad:2013xma,Chatrchyan:2012wg,CMS:2012zex,Chatrchyan:2012xq}. Comparison to theoretical model calculations have shown that the QGP behaves like an almost perfect fluid \cite{Heinz:2013th,Luzum:2013yya,Shuryak:2014zxa,Song:2017wtw} with very low specific shear viscosity, $\eta/s$. A good theoretical model description on the experimental data have a significant impact on the understanding of the QGP. The current state-of-the-art understanding of the QGP comes from models based on Bayesian inference \cite{Bernhard:2019bmu,Everett:2020xug,Nijs:2020roc}, which use multiple experimental measurements to simultaneously constrain the parameters of the initial conditions and the hydrodynamical model. The extracted information can be better constrained by including higher-order flow and correlations of different order flow harmonics in the experimental observables \cite{Parkkila:2021yha}. Additional constraints can be obtained by including the \pt-differential studies of the flow coefficient $v_n(\pt)$ and its fluctuations, as they contain additional information about the dynamics of the expansion and the initial conditions of the heavy-ion collisions \cite{Nijs:2020roc}. Fluctuations of the flow vector with the transverse momentum, \pt, are predicted in hydrodynamical simulations \cite{Heinz:2013bua,Gardim:2012im} and have been measured at the LHC \cite{ALICE:2017lyf,CMS:2013bza,Khachatryan:2015oea,Aamodt:2011by}. The event-by-event fluctuations of the flow coefficient $v_n$ are well known \cite{Voloshin:2008dg,Heinz:2013th,ALICE:2022zks}; however, the decorrelation of the flow coefficient in \pt has not been studied separately from the flow angle fluctuations \cite{ALICE:2017lyf}. Additionally, the fluctuations of the flow angle $\Psi_n$ has proved hard to measure experimentally \cite{Qiu:2011iv,Teaney:2010vd,Jia:2012ma,Aad:2014fla,Qiu:2012uy,Gardim:2012im,Bozek:2018nne}, and the dependence of the flow angle on the transverse momentum \cite{Gardim:2012im,Gardim:2017ruc,Zhao:2017yhj,Bozek:2018nne,Barbosa:2021ccw} results in a decorrelation between the integrated symmetry plane, $\Psi_n$, and the \pt-differential plane, $\Psi_n(\pt)$. Such fluctuations potentially affect measurements that rely on the assumption of a common symmetry plane. The overall flow vector fluctuation gives insight into the event-by-event fluctuating initial conditions and separating the fluctuations into the flow angle and flow magnitude fluctuations enables the individual study of the sources of the fluctuations in heavy-ion collisions. Measurements of \pt-dependent flow vector fluctuations have traditionally been measured with two-particle correlations \cite{Heinz:2013bua,Gardim:2012im,ALICE:2017lyf,CMS:2013bza,Khachatryan:2015oea}. Recently, we proposed new four-particle correlations to measure the flow angle and flow magnitude fluctuations separately \cite{IS2021}, which cannot be done with two-particle correlations. A follow-up study confirmed that the flow angle and magnitude fluctuation phenomena were present in hydrodynamical models \cite{Bozek:2018nne,Bozek:2021mov}. Additional hydrodynamical model calculations are compared to the ALICE data in \cite{ALICE:2022smy}, which were successful in capturing the trend of the data. However, the model calculations of the flow magnitude fluctuations showed a puzzling trend when using large values of $\eta/s$. At low \pt, where the expectation is to see little to no fluctuations, these calculations show values of the flow magnitude fluctuations greatly above 1. Meanwhile, systematic studies with a transport model are completely missing. Thus, we present the first transport model study on this topic. \\
In this paper, the \pt-dependent flow vector, flow angle and flow magnitude fluctuations are studied with A Multi-Phase Transport model (AMPT) \cite{Lin:2004en,Lin:2021mdn}. The AMPT model is well suited to the study of anisotropic flow observables. It reproduces measurements of anisotropic flow at RHIC \cite{Lin:2001zk,Xu:2011fe} and the LHC \cite{Xu:2011fi,Feng:2016emh}. The analysis in this work is performed with different tunings of the model in order to explore the effects of the initial conditions, transport properties of the QGP and the hadronic rescatterings. In section \ref{sec:level2}, details of the model and the different configurations are presented. Section \ref{sec:level3} describes the observables and the method used to extract information about the decorrelation of the flow vector. The results are presented and discussed in section \ref{sec:level4}. Finally, the work is summarized in section \ref{sec:level5}. Ratio plots between the different AMPT configurations for all observables can be found in appendix \ref{app:ratios}.
\section{\label{sec:level2}The model}
In this paper, AMPT \cite{Lin:2004en,Lin:2021mdn} is used to study the transverse momentum decorrelation of the flow vector, flow angle and flow magnitude in \PbPb collisions at 5.02 TeV. 
The model consists of four stages: initial conditions, partonic scattering, hadronization, and hadronic rescattering. The initial conditions are generated with HIJING \cite{PhysRevD.44.3501}, which includes the spatial and momentum distributions of mini-jet partons and soft string excitations. Hadrons produced from string fragmentation are then converted with string melting into their valence quarks and anti-quarks. The interaction of partons is treated with ZDC \cite{Zhang:1997ej}, which only includes two-body scatterings. The cross sections are obtained from pQCD with screening masses
\begin{align}
\sigma=\frac{9\pi\alpha_s^2}{2\mu^2},
\label{eq:Xsec}
\end{align}
where $\alpha_s$ is the strong coupling and $\mu$ is the Debye screening mass. At LHC energies, a strong coupling constant of $\alpha_s = 0.33$ is used \cite{Lin:2014tya}. Partons are converted into hadrons with a quark coalescence model, combining two quarks into mesons and three quarks into baryons \cite{Chen:2005mr}. Finally, the hadronic rescattering is handled by ART \cite{Li:1995pra,Li:2001xh}, where the dynamics of the hadronic matter are described by a hadron cascade. In order to study the effects of the initial state on the flow vector fluctuations, the Lund string parameters a and b, which are related to the string tension, are varied. The default setting is $a=0.3$ and $b=0.15$ \cite{Lin:2014tya,Feng:2016emh}, and the variation is $a=0.5$ and $b=0.9$, which are the default HIJING values \cite{Wang:1991hta,Gyulassy:1994ew}. The effect of the QGP properties is studied by varying the partonic cross section using different values of the screening mass $\mu$. The values $\mu = 3.2042~\mathrm{fm}^{-1}$ and $\mu = 2.2814~\mathrm{fm}^{-1}$ correspond to cross sections of $\sigma\approx1.5$ mb and $\sigma\approx3$ mb, respectively. A smaller cross section in AMPT corresponds to a larger viscosity in viscous hydrodynamics \cite{Gyulassy:1997ib,Zhang:1999rs}. The specific shear viscosity can be approximated by \cite{Xu:2011fi}
\begin{align}
\eta/s \approx \frac{3\pi}{40\alpha_s^2}\frac{1}{\left(9+\frac{\mu^2}{T^2}\right)ln\left(\frac{18+\mu^2/T^2}{\mu^2/T^2}\right)-81},
\end{align}
where $T$ is the initial temperature. At the LHC, the initial temperature is approximately $468$ MeV \cite{Xu:2011fi}, which leads to values of specific shear viscosity around $0.18$ for 3 mb and $0.28$ for 1.5 mb. Additionally, the effects of the hadronic rescatterings are explored by changing the rescattering time in ART between 0.6 fm/\textit{c} (ART effectively OFF) and 30 fm/\textit{c} (ART ON). In total, five different sets of parameters are used for AMPT in this study, as seen in table \ref{tab:Pars}. 
The results in this paper are presented in different centrality intervals. The centrality is determined by the impact parameter \textit{b} as in \cite{Xu:2011fi}. Around 320K events are generated in the 0-10\% central regions, and around 300K events are generated for each 10\% interval of the other centralities.
\begin{center}
\begin{table}[h!]
\begin{tabular}{c|c|c|c|c}
	&Partonic cross section	&a	&b &ART\\\hline
Par1	&1.5 mb	&0.3	&0.15	&ON\\\hline
Par2	&1.5 mb	&0.3	&0.15	&OFF\\\hline
Par3	&1. 5mb	&0.5	&0.9	&OFF\\\hline		
Par4	&3.0 mb	&0.3	&0.15	&ON\\\hline
Par5	&3.0 mb	&0.3	&0.15	&OFF\\\hline
\end{tabular}
\caption{Table of the different sets of parameters used for AMPT in this study. The cross section is determined from the screening mass $\mu$ with Eq. (\ref{eq:Xsec}).}
\label{tab:Pars}
\end{table}
\end{center}
\section{\label{sec:level3}Observables}
The \pt-dependent flow vector fluctuations are studied with \pt-differential flow coefficients. The regular \pt-differential flow coefficient is defined as
\begin{align}
v_n\{2\} &= \frac{\la\la\cos n(\varphi_1^\mathrm{POI}-\varphi_2)\ra\ra}{\sqrt{\la\la\cos n(\varphi_1-\varphi_2)\ra\ra}}\nonumber\\
&=\frac{\la v_n(\pt)v_n\cos n[\Psi_n(\pt)-\Psi_n]\ra}{\sqrt{\la v_n^2\ra}},\label{eq:vnDiff}
\end{align}
where $\varphi^\mathrm{POI}$ is the azimuthal angle of a \textit{particle of interest} (POI) chosen from a narrow \pt range and $\varphi$ is the azimuthal angle of a reference particle (RP) chosen from a wide \pt range. The single bracket denotes an average over all events, and double brackets indicate an average over all particles and all events. The $v_2\{2\}$ probes the \pt-differential flow and is sensitive to fluctuations of both the flow magnitude and flow angle. Another \pt-differential flow observable was proposed in \cite{Heinz:2013bua}, which is unaffected by the flow vector fluctuations
\begin{align}
v_n[2] &= \sqrt{\la\la\cos n(\varphi_1^\mathrm{POI}-\varphi_2^\mathrm{POI})\ra\ra}\nonumber\\
&= \sqrt{\la v_n^2(\pt)\ra}\label{eq:vnPtA}.
\end{align}
Since $v_n[2]$ is not affected by the flow angle and flow magnitude fluctuations, the \pt-dependent flow vector fluctuations can be probed by taking the ratio of $v_n\{2\}$ and $v_n[2]$: 
\begin{align}
\frac{v_n\{2\}}{v_n[2]} = \frac{\la v_n(\pt)v_n\cos n[\Psi_n(\pt)-\Psi_n]\ra}{\sqrt{\la v_n^2(\pt)\ra}\sqrt{\la v_n^2\ra}}\label{eq:vnratio}.
\end{align}
A value of $v_n\{2\}/v_n[2]<1$ indicates the presence of \pt-dependent flow vector fluctuations. Similarly, the double-differential factorization ratio \cite{Gardim:2012im}, which tests the factorization of the two-particle correlation into the product of single-particle flow coefficients, is a measure of the flow vector fluctuations. It provides more detailed information about the correlation structure
\begin{align}
r_n &= \frac{\la\la\cos n(\varphi_1^a-\varphi_2^t)\ra\ra}{\sqrt{\la\la\cos n(\varphi_1^a-\varphi_2^a)\ra\ra\la\la\cos n(\varphi_1^t-\varphi_2^t)\ra\ra}}\nonumber\\
&= \frac{\la v_n(\pta)v_n(\ptt)\cos n[\Psi_n(\pta)-\Psi_n(\ptt)]\ra}{\sqrt{\la v_n^2(\pta)\ra}\sqrt{\la v_n^2(\ptt)\ra}}\label{eq:rn}.
\end{align}
Here, $\varphi_2^a$ is the azimuthal angle of the \textit{associate} particle, and $\varphi_2^t$ is the azimuthal angle of the \textit{trigger} particle with transverse momentum \pta and \ptt, respectively. The \ptt-range is fixed for all values of \pta. Again, a value of $r_n<1$ indicates that factorization is broken and that \pt-dependent flow vector fluctuations are present. Both Eqs. (\ref{eq:vnratio}) and (\ref{eq:rn}) are based on two-particle correlations and may be affected by fluctuations in both the flow magnitude and flow angle. Previously we proposed to use four-particle correlations to separate the two types of fluctuations \cite{IS2021} so that they can be measured experimentally. The flow angle fluctuations are given by
\begin{align}
A_n^\mathrm{f} &= \frac{\la \cos n(\varphi_1^a+\varphi_2^a-\varphi_3-\varphi_4)\ra}{\la\cos n(\varphi_1^a+\varphi_2-\varphi_3^a-\varphi_4)\ra}\nonumber\\
&= \frac{\la v_n^2(\pt)v_n^2\cos 2n[\Psi_n(\pt)-\Psi_n]\ra}{\la v_n^2(\pt)v_n^2\ra}\label{eq:Af}\\
&\approx \la\cos 2n[\Psi_n(\pt)-\Psi_n]\ra,\nonumber
\end{align}
where the last equality holds if the non-flow is approximately the same in numerator and denominator. Then $A^\mathrm{f}_n<1$ suggests \pt-dependent flow angle fluctuations. The flow magnitude is given by the double ratio of the \pt-differential flow fluctuations against the reference flow baseline fluctuations
\begin{align}
M_n^\mathrm{f} &= [\la\la \cos n[\varphi_1^\mathrm{a}+\varphi_2-\varphi_3^\mathrm{a}-\varphi_4]\ra\ra /\nonumber\\
&\left(\la\la \cos n[\varphi_1^\mathrm{a}-\varphi_3^\mathrm{a}]\ra\ra\la\la\cos n[\varphi_2-\varphi_4]\ra\ra\right)]/\nonumber\\
&[\la\la\cos n[\varphi_1+\varphi_2-\varphi_3-\varphi_4]\ra\ra/\la\la \cos n[\varphi_1-\varphi_2]\ra\ra^2]\nonumber\\
&= \frac{\la v_n^2(\pt)v_n^2\ra/(\la v_n^2(\pt)\ra\la v_n^2\ra}{\la v_n^4\ra/\la v_n^2\ra^2}
\label{eq:Mf}
\end{align}
Any deviation from unity indicates the presence of \pt-dependent flow magnitude fluctuations.

\section{\label{sec:level4}Method}
The anisotropic flow observables are calculated using the two- and multi-particle correlation method \cite{Borghini:2001vi,Bilandzic:2010jr}. This method has been widely used at RHIC \cite{STAR:2011ert} and the LHC \cite{Aamodt:2010pa,ALICE:2011ab} in the study of anisotropic flow. In this study, the Generic Framework \cite{Bilandzic:2013kga,Huo:2017nms} is used, which enables a fast and exact calculation of the multi-particle correlations. However, additional correlators had to be implemented in the Generic Framework to separate the flow angle and flow magnitude fluctuations as in Eqs. (\ref{eq:Af}) and (\ref{eq:Mf}). While the method has been used for the measurements in \cite{ALICE:2022smy}, a detailed explanation of the method has not been available before this paper.
\subsection{Standard multi-particle correlations}
The anisotropic flow correlations are expressed in terms of the Q-vector
\begin{align}
Q_{n,p} = \sum_{k=1}^Mw_k^pe^{in\varphi_k},
\label{eq:Qvec}
\end{align}
where $M$ is the multiplicity of the selected particles in a given kinematic range, $\varphi$ is the azimuthal angle of the particles, and $w$ is a particle weight that corrects for detector inefficiencies in experimental measurements. In general, the single-event m-particle correlation can be calculated with
\begin{align}
\la m\ra_{n_1,\ldots,n_m} = \frac{\displaystyle\sum_{k_1\neq\ldots\neq k_m}^M w_{k_1}\ldots w_{k_m}e^{i(n_1\varphi_1+\ldots +n_m\varphi_m)}}{\displaystyle\sum_{k_1\neq\ldots\neq k_m}^M w_{k_1}\ldots w_{k_m}},\label{eq:mcorr}
\end{align}
The numerator and denominator in Eq. (\ref{eq:mcorr}) are trivially related
\begin{align}
\textrm{N}\la m\ra_{n_1,\ldots,n_m} &= \displaystyle\sum_{k_1\neq\ldots\neq k_m}^M w_{k_1}\ldots w_{k_m}e^{i(n_1\varphi_1+\ldots +n_m\varphi_m)}\\
\textrm{D}\la m\ra_{n_1,\ldots,n_m} &= \displaystyle\sum_{k_1\neq\ldots\neq k_m}^M w_{k_1}\ldots w_{k_m} \\
&= \textrm{N}\la m\ra_{0,\ldots,0}\nonumber
\end{align}
Due to the trivial nature of the denominators, they will be omitted from the following part unless necessary. The exact equations for the two- and four-particle azimuthal single-event correlations $\mathrm{N}\la 2\ra_{n_1,n_2}$ and $\mathrm{N}\la 4\ra_{n_1,n_2,n_3,n_4}$ calculated with the Generic Framework are given by Eqs. (19)-(22) in \cite{Bilandzic:2013kga}. The event-averaged two- and four-particle correlations can then be obtained with
\begin{align}
\la\la 2\ra\ra &= \frac{\displaystyle\sum_{\mathrm{events}}\mathrm{N}\la 2\ra_{n_1,n_2}}{\displaystyle\sum_{\mathrm{events}}\mathrm{D}\la 2\ra_{n_1,n2}},\label{eq:evcorr2}\\
\la\la 4\ra\ra &= \frac{\displaystyle\sum_{\mathrm{events}}\mathrm{N}\la 4\ra_{n_1,n_2,n_3,n_4}}{\displaystyle\sum_{\mathrm{events}}\mathrm{D}\la 4\ra_{n_1,n_2,n_3,n_4}},\label{eq:evcorr4}
\end{align}
where the double brackets refer to an average over all particles and events. $\mathrm{D}\la m\ra_{n_1,n_2,\ldots,n_m}$ is used as an event weight in order to reduce the effect of the varying multiplicity \cite{Bilandzic:2013kga}. 

For \pt-differential anisotropic flow, one or more of the particles are selected from a narrow \pt range by constructing a p-vector analogous to the Q-vector consisting only of particles of interest.
\begin{align}
p_{n,p} = \sum_{k=1}^{M_\mathrm{POI}}w_k^pe^{in\varphi}\label{eq:pVec},
\end{align}
where $M_\mathrm{POI}$ is the multiplicity of the POIs in the selected kinematic region. The p-vector enables the calculation of various \pt-differential anisotropic flow coefficients, such as $v_n\{2\}$ by selecting one POI and one RP or $v_n[2]$ by selecting two POIs, in the calculation of the two-particle correlations. Additionally, the double-differential two-particle correlations are calculated with one \textit{associate} particle and one \textit{trigger} particle from different narrow \pt-ranges. The \pt-differential two-particle correlations are then
\begin{align}
\mathrm{N}\la 2'\ra_{n_1,n_2} &= Q_{n_1,1}p_{n_2,1}-q_{n_1+n_2,2},\label{eq:diff2}\\
\mathrm{N}\la 2''\ra_{n_1,n_2} &= p_{n_1,1}p_{n_2,1}-p_{n_1+n_2,2},\label{eq:pta2}\\
\mathrm{N}\la 2''\ra_{n_1,n_2}^{a,t} &= p_{n_1,1}^ap_{n_2,1}^t,\label{eq:ptaptt2}
\end{align}
where $q_{n,p}$ is the overlap-vector of particles that are both POIs and RPs, and the superscript $a$ refers to \textit{associate} particles, and $t$ refers to \textit{trigger} particles. Eq. (\ref{eq:diff2}) takes one POI and correlates it with the reference particles, whereas Eq. (\ref{eq:pta2}) correlates two POIs from the same \pt-range. Eq. (\ref{eq:ptaptt2}) is a special case of Eq. (\ref{eq:pta2}) when the two POIs are taken from different \pt-ranges. The event-averaged \pt-differential correlations then follow similarly to Eqs. (\ref{eq:evcorr2}) and (\ref{eq:evcorr4})
\begin{align}
\la\la 2'\ra\ra &= \frac{\displaystyle\sum_{\mathrm{events}}\mathrm{N}\la 2'\ra_{n_1,n_2}}{\displaystyle\sum_{\mathrm{events}}\mathrm{D}\la 2'\ra_{n_1,n2}},\label{eq:evcorr2p}\\
\la\la 2''\ra\ra &= \frac{\displaystyle\sum_{\mathrm{events}}\mathrm{N}\la 2''\ra_{n_1,n_2}}{\displaystyle\sum_{\mathrm{events}}\mathrm{D}\la 2''\ra_{n_1,n2}},\label{eq:evcorr2pp}\\
\la\la 2'\ra\ra &= \frac{\displaystyle\sum_{\mathrm{events}}\mathrm{N}\la 2\ra_{n_1,n_2}^{a,t}}{\displaystyle\sum_{\mathrm{events}}\mathrm{D}\la 2\ra_{n_1,n2}^{a,t}},\label{eq:evcorr2at}
\end{align}
\subsection{Subevent method}
Non-flow effects are correlations not due to collective effects, such as resonance decays and jets. They can be suppressed by applying a gap in pseudorapidity, $|\Delta\eta|$, between two subevents, A and B, from which the particles are selected.
\begin{align}
\mathrm{N}\la 2\ra_{n1,n2}^{|\Delta\eta|} &= Q_{n_1,1}^A\cdot Q_{n_2,1}^{B},\label{eq:2gap}\\
\mathrm{N}\la 4\ra_{n_1,n_2,n_3,n_4}^{|\Delta\eta|} &= \mathrm{N}\la 2\ra_{n_1,n_2}^A\cdot\mathrm{N}\la 2\ra_{n_1,n_2}^B.\label{4gap}
\end{align}
For \pt-differential correlators the subevents are implemented analogous to Eq. (\ref{eq:2gap})
\begin{align}
\mathrm{N}\la 2'\ra_{n_1,n_2}^{|\Delta\eta|} = p_{n_1,1}^AQ_{n_2,1}^{B},\label{eq:diff2gap}\\
\mathrm{N}\la 2''\ra_{n_1,n_2}^{|\Delta\eta|} = p_{n_1,1}^Ap_{n_2,1}^{B},\label{eq:pta2gap}\\
\mathrm{N}\la 2''\ra_{n_1,n_2}^{a,t,|\Delta\eta|} = p_{n_1,1}^{a,A}p_{n_2,1}^{t,B}.\label{eq:ptaptt2gap}
\end{align}
\begin{figure}[t]
\centering
\includegraphics[scale=0.45]{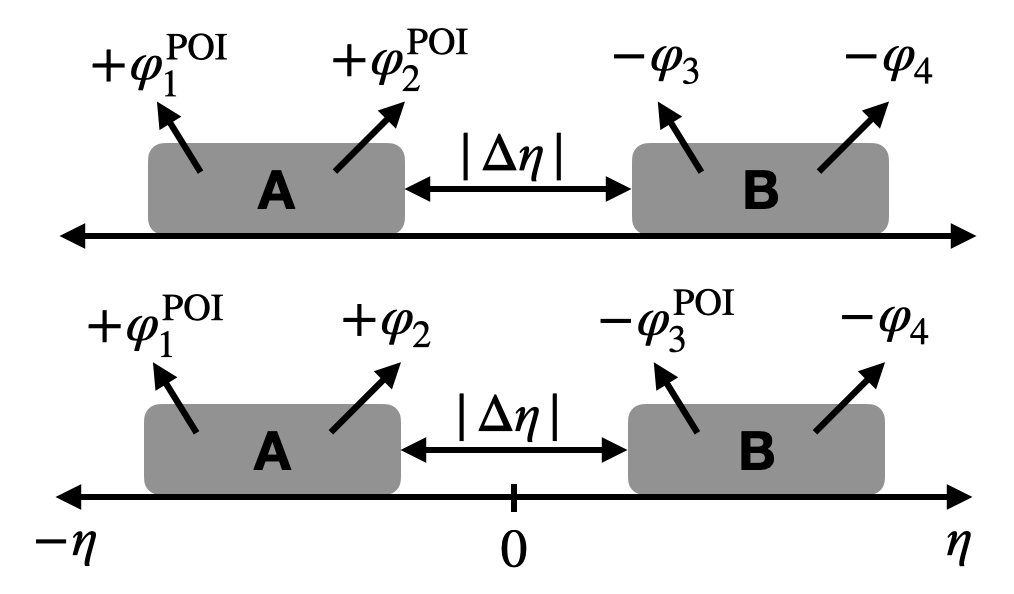}
\caption{\label{fig:Subevents}The event divided into two subevents separated by a gap in pseudorapidity with two POIs selected from the same subevent (top) and one POI selected from each subevent (bottom).}
\end{figure}
The four-particle correlations, which are needed to calculate Eq. (\ref{eq:Af}) and (\ref{eq:Mf}), are obtained with the subevent method to reduce the possible non-flow contamination. For four-particle correlations with a pseudorapidity gap, there is a choice of how to group the POIs and RPs. Either the POIs are chosen from the same subevents (SS), or one POI is chosen from opposite subevents (OS), as shown in fig. \ref{fig:Subevents}. 
\begin{align}
\mathrm{N}\la 4'\ra_\mathrm{SS} = &(p_{n_1,1}p_{n_2,1}-p_{n_1+n_2,2})^A\nonumber\\&\cdot(Q_{n_3,1}Q_{n_4,1}-Q_{n_3+n_4,2})^{B},\label{4SS}\\
\mathrm{N}\la 4'\ra_\mathrm{OS} = &(p_{n_1,1}Q_{n_2,1}-q_{n_1+n_2,2})^A\nonumber\\&\cdot(p_{n_3,1}Q_{n_4,1}-q_{n_3+n_4,2})^{B}.\label{eq:4OS}
\end{align}
The event average of the correlators described in Eqs. (\ref{eq:diff2})-(\ref{eq:4OS}) are obtained by replacing the correlators in Eqs. (\ref{eq:evcorr2}) and (\ref{eq:evcorr4}). The four-particle correlations are, by construction, less sensitive to non-flow correlations, as the latter typically involves fewer particles. By also applying a pseudorapidity gap, it ensures that the non-flow contamination is minimal. The non-flow can also be tested by performing the analysis with only particles with the same sign charge, also known as like-sign method. It was found in \cite{ALICE:2022smy} that the method is robust against non-flow within 1\% from the like-sign method as well as from model studies with HIJING, which only contains non-flow correlations. The four-particle correlations presented here are identical to the method used in the ALICE measurements. This is, however, different to the four-particle correlations used in \cite{Magdy:2022jai}, where short-range correlations between particles in each subevents are present. These can be significantly biased by non-flow and thus lead to a different conclusion. \\

With the above definitions, the observables to probe the \pt-dependent flow vector fluctuations can easily be calculated. The \pt-differential flow coefficient $v_n\{2\}$ is
\begin{align}
v_n\{2\} &=\frac{\la\la 2'\ra\ra}{\sqrt{\la\la 2\ra\ra}},\label{eq:vnDiffGF}
\end{align}
and the alternative \pt-differential flow coefficient $v_n[2]$ becomes
\begin{align}
v_n[2] &= \sqrt{\la\la 2''\ra\ra}.\label{eq:vnPtAGFW}
\end{align}
The factorization ratio $r_n$ is given by
\begin{align}
r_n &= \frac{\la\la 2''\ra\ra_{a,t}}{\sqrt{\la\la 2''\ra\ra_a\la\la 2''\ra\ra_t}}.\label{eq:rnGFW}
\end{align}
Finally, the observables based on four-particle correlations, $A_n^\mathrm{f}$ and $M_n^\mathrm{f}$, which describe the flow angle and flow magnitude fluctuations, respectively, are given by
\begin{align}
A_n^\mathrm{f} &= \frac{\la\la 4'\ra\ra_\mathrm{SS}}{\la\la 4'\ra\ra_\mathrm{OS}},\label{eq:AfGFW}\\
M_n^\mathrm{f} &= \frac{\la\la 4'\ra\ra_\mathrm{OS}/(\la\la 2''\ra\ra\la\la 2\ra\ra)}{\la\la 4\ra\ra/\la\la 2\ra\ra^2}.\label{eq:MfGFW}
\end{align}
\begin{figure*}[t]
\centering
\includegraphics[scale=0.9]{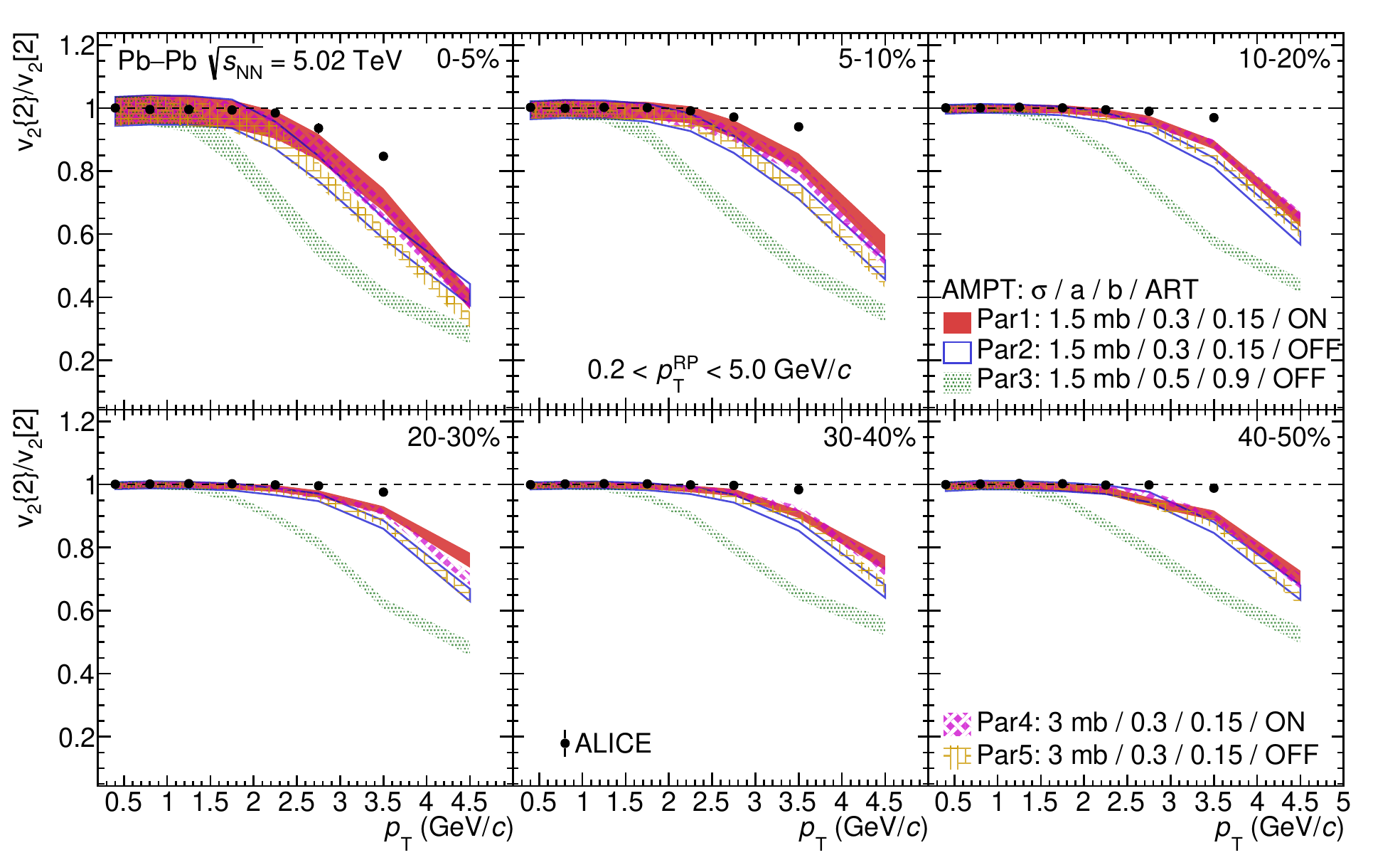}
\caption{\label{fig:v2ratio}The \pt-dependence of $v_2\{2\}/v_2[2]$ in different centrality classes from the AMPT model with different configurations: $\sigma=$ 1.5 mb and ART ON (Par1), $\sigma=$ 1.5 mb and ART OFF (Par2), $\sigma=$1.5 mb, ART ON, $a=0.5$ and $b=0.9$ (Par3), $\sigma=$ 3.0 mb and ART ON (Par4), and $\sigma=$ 3.0 mb and ART OFF (Par5). ALICE data points are shown as black circles for comparison.}
\end{figure*}
As the non-flow contribution is small for multi-particle cumulants \cite{ALICE:2019zfl}, a smaller $\eta$-gap is used for the four-particle correlations compared to the two-particle correlations to increase the event sample size. No particle weights are used to correct detector inefficiencies and acceptance for the generated AMPT events. \\
\section{\label{sec:level5}Results and discussion}
\begin{figure*}
\centering
\includegraphics[scale=0.92]{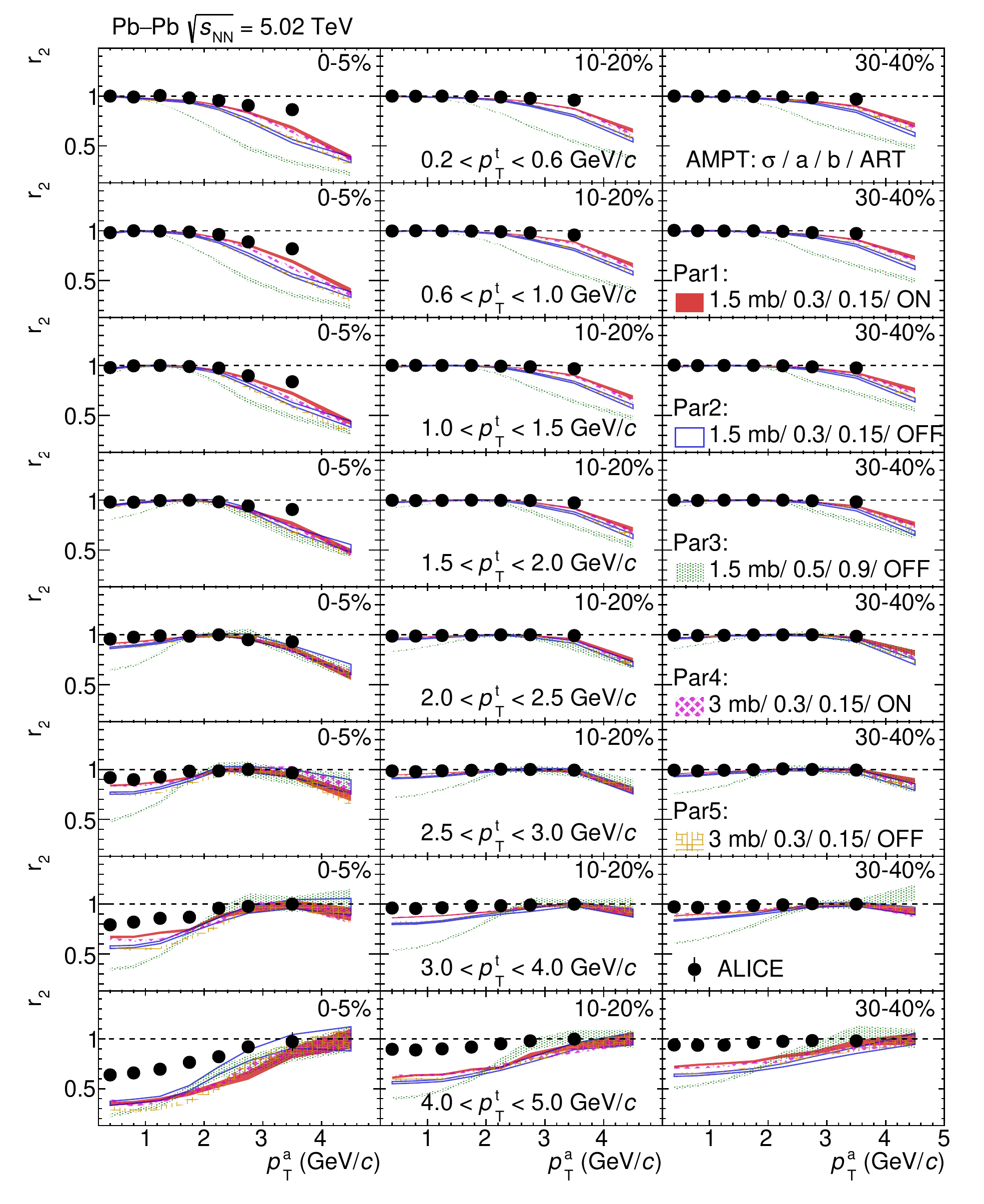}
\caption{\label{fig:r2}The \pt-dependence of the factorization ratio $r_2$ in different centrality and \textit{trigger} particle transverse momentum ranges from the AMPT model with different configurations: $\sigma=$1.5 mb and ART ON (Par1), $\sigma=$1.5 mb and ART OFF (Par2), $\sigma=$1.5 mb, ART ON, $a=0.5$ and $b=0.9$ (Par3), $\sigma=$3.0 mb and ART ON (Par4), and $\sigma=$3.0 mb and ART OFF (Par5). ALICE data points are shown as black circles for comparison.}
\end{figure*}
\begin{figure*}[t]
\centering
\includegraphics[scale=0.8]{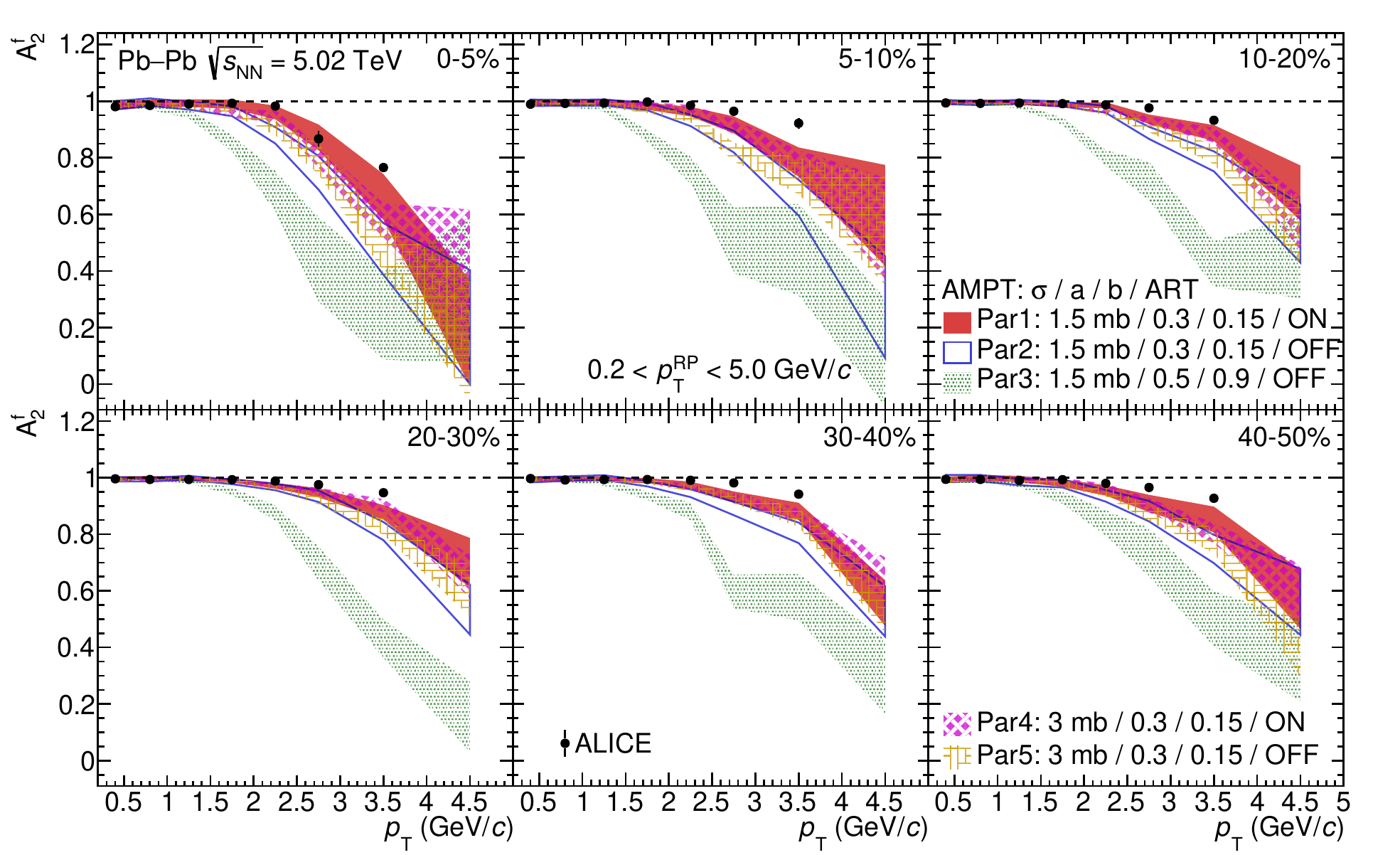}
\caption{\label{fig:A2}The \pt-dependence of the flow angle fluctuations, $A_2^\mathrm{f}\approx \la\cos 2n[\Psi_n(\pt)-\Psi_n]\ra$, in different centrality ranges from the AMPT model with different configurations: $\sigma=$1.5 mb and ART ON (Par1), $\sigma=$1.5 mb and ART OFF (Par2), $\sigma=$1.5 mb, ART ON, $a=0.5$ and $b=0.9$ (Par3), $\sigma=$3.0 mb and ART ON (Par4), and $\sigma=$3.0 mb and ART OFF (Par5). ALICE data points are shown as black circles for comparison \cite{ALICE:2022smy}.}
\end{figure*}
The \pt-dependence of the $v_2\{2\}/v_2[2]$ ratio calculated from the different configurations of the AMPT model is shown in fig. \ref{fig:v2ratio} for centrality ranges 0-5\% to 40-50\%. The deviation of $v_2\{2\}/v_2[2]$ from unity is largest in the 0-5\% central collisions and then becomes progressively smaller as the centrality increases. The effect of changing the initial conditions via the Lund string parameters, a and b, can be seen by comparing the results from Par3 and Par2. The effect depends on the transverse momentum with an increasing effect up to 4 GeV/\textit{c}, after which it decreases slightly. The string parameters affect the ratio $v_2\{2\}/v_2[2]$ by up to 35\% in central collisions and up to 30\% in non-central collisions. The $v_2\{2\}/v_2[2]$ shows a weak dependence on the partonic cross section: comparing the different cross sections with ART ON shows at most a 10\% difference (Par1 vs Par4). Without ART, the effect of changing the cross section is negligible except in the 0-5\% central collisions (Par2 vs Par5). The comparison with ART OFF may better reflect the true sensitivity to the QGP properties as the final state hadronic interactions may differ significantly when ART is enabled (Par1 vs Par2, Par4 vs Par5). A smaller partonic cross section in AMPT corresponds to a larger specific viscosity in hydrodynamics. The expectation is therefore that the calculation of $v_2\{2\}/v_2[2]$ with a smaller partonic cross section should show less deviation from unity as a larger specific shear viscosity tends to dampen fluctuations from the initial state \cite{Schenke:2010rr}. This expectation is consistent with what is observed in fig. \ref{fig:v2ratio}: the Par1 calculation with smaller partonic cross section deviates less from unity than the Par4 calculations. However, the difference is small compared to the sensitivity to the initial state. The effect of the hadronic rescattering on the ratio $v_2\{2\}/v_2[2]$ increases with \pt and is at most 10\% in the presented \pt-range across all centralities for the Par5 vs Par4 calculations. The same is true for the Par2 vs Par1 results except in the 0-5\% most central collisions, where a 20\% difference is seen in the highest \pt bin. The above comparisons suggest that the flow vector fluctuations are highly sensitive to the initial state of heavy-ion collision and, unlike the $v_2$ flow coefficient itself, are less sensitive to the transport properties of the created matter. The effect of the hadronic rescattering (ART ON vs ART OFF) is constant across the non-central collisions. Measurements of $v_2\{2\}/v_2[2]$ from the ALICE experiment in \PbPb collisions at \snn = 5.02 TeV are shown for comparison. All the configurations of the AMPT model overestimate the deviation from unity, especially in non-central collisions, suggesting that the model needs further tuning to the data, particularly of the initial conditions. The calculation with 1.5 mb cross section, a = 0.3, b = 0.15 and ART ON (Par1) is closest in describing the data and only overestimates the data in the highest \pt data point.

The factorization ratio of the second-order flow harmonic, $r_2$, is shown in fig. \ref{fig:r2} as a function of \textit{associate} particle transverse momentum, \pta, in centrality ranges 0-5\%, 10-20\% and 30-40\% and different \textit{trigger} particle transverse momentum, \ptt, ranges. The factorization ratio in centrality ranges 5-10\%, 20-30\% and 40-50\% is shown in Appendix \ref{app:rn}. The deviation of the factorization ratio from unity is largest in the 0-5\% central collisions and increases with the difference $|\pta-\ptt|$, which has also been observed in both hydrodynamics \cite{Gardim:2012im} and data \cite{Acharya:2017ino,Khachatryan:2015oea}. The different configurations of AMPT all predict factorization breaking, with the largest deviation from unity observed for the Par3 model, which has different parameters for the initial conditions. The effect of the initial stages on the factorization ratio by changing the Lund string parameters from $a=0.3$, $b = 0.15$ to $a = 0.5$, $b = 0.9$ is up to 40\% in central collisions and increases with $|\pta-\ptt|$. However, for the lower \ptt-ranges, $\ptt<1.5$ GeV/\textit{c}, the effect does not increase for $\pta > 3$ GeV/\textit{c} for central collisions and $\pta > 4$ GeV/\textit{c} for non-central collisions. In non-central collisions, the effect is up to 30\% for large values of $|\pta-\ptt|$. Changing the cross section from 1.5 mb to 3.0 mb affects $r_2$ by up to 10\% at high $|\pta-\ptt|$ with ART OFF (Par5 vs Par2). This is consistent with $v_2\{2\}/v_2[2]$, as expected since $r_2$ is basically the double-differential $v_2\{2\}/v_2[2]$. The hadronic phase also shows an effect of around 10\% at high $|\pta-\ptt|$ across all centralities for 1.5 mb (Par2 vs Par1) and 3.0 mb (Par5 vs Par4). The ALICE measurements favour the model with a 1.5 mb cross section, a = 0.3, b = 0.15 and ART ON (Par1), although the model overestimates the breaking of factorization as $|\pta-\ptt|$ increases. 

\begin{figure*}
\centering
\includegraphics[scale=0.8]{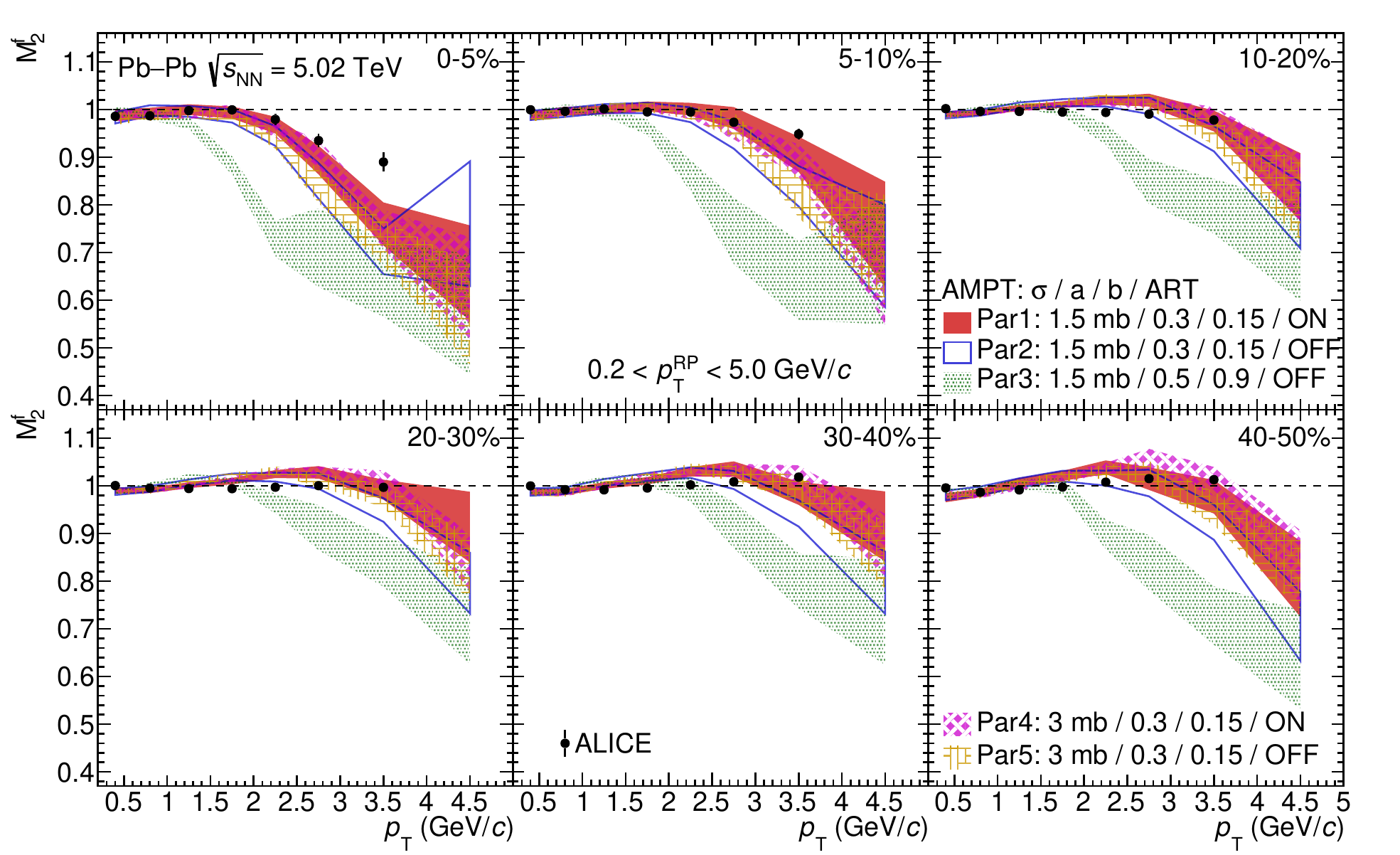}
\caption{\label{fig:M2}The \pt-dependence of the flow magnitude fluctuations, $M_2^\mathrm{f}$, in different centrality ranges from the AMPT model with different configurations: $\sigma=$1.5 mb and ART ON (Par1), $\sigma=$1.5 mb and ART OFF (Par2), $\sigma=$1.5 mb, ART ON, $a=0.5$ and $b=0.9$ (Par3), $\sigma=$3.0 mb and ART ON (Par4), and $\sigma=$3.0 mb and ART OFF (Par5). ALICE data points are shown as black circles for comparison \cite{ALICE:2022smy}.}
\end{figure*}
In fig. \ref{fig:A2}, the flow angle fluctuations $A_2^\mathrm{f}$ are shown as a function of transverse momentum in centrality ranges 0-5\% to 40-50\%. A deviation of $A_2^\mathrm{f}$ from unity is observed for all the model calculations, with the majority showing deviations for $\pt > 2.5$ GeV/\textit{c} and the Par3 calculation with $a=0.5,b=0.9$ showing deviation from unity for $\pt>1.5$ GeV/\textit{c}. The deviation from unity indicates that the \pt-dependent flow angle, $\Psi_n(\pt)$ fluctuates around the \pt-integrated flow angle $\Psi_n$ in the AMPT model, and the fluctuations are strongest in the central collisions, where the initial state density fluctuations dominate. However, significant fluctuations of up to 50\% for Par3 and 30-40\% for the other configurations can be seen across the centralities.
The flow angle fluctuations are highly sensitive to the initial conditions. Changing the Lund string parameters for the initial conditions affects the $\Fpsi{2}$ by up to 70\% in the 20-30\% centrality range as shown in \ref{fig:A2} bottom left (Par3 vs Par2). The large uncertainties at high \pt in central collisions do not allow for a quantitative statement about the effect, but at lower \pt, the effect is up to 50\%. The flow angle fluctuations show no sensitivity to the partonic cross section within the uncertainties whether ART is ON (Par4 vs Par1) or OFF (Par5 vs Par2). No significant effect of the hadronic rescattering is observed at high \pt due to the large uncertainties. At lower \pt, the effect is $\sim20\%$ for the 1.5 mb calculations (Par2 vs Par1) and even less for the 3.0 mb calculations (Par5 vs Par4). Therefore, the fluctuating flow angles are most likely driven mainly by the event-by-event fluctuations in the initial state. The AMPT model with 1.5 mb cross section, Lund string parameters $a=0.3$ and $b=0.15$, and with hadronic interactions (Par1) shows the most accurate description of the ALICE measurements, only slightly overestimating the deviation from unity at $\pt > 3.0$ GeV/\textit{c} in the 0-5\% and 5-10\% central collisions and at $\pt > 2.0$ GeV/\textit{c} in centralities 30-40\% and 40-50\%. The flow angle fluctuations, which have been observed in data and hydrodynamic models, are also reproduced in the AMPT model.

The flow magnitude fluctuations $M_2^\mathrm{f}$ are shown in fig. \ref{fig:M2} as a function of transverse momentum in centrality ranges 0-5\% to 40-50\%. As $M_2^\mathrm{f}$ is normalized to the baseline \pt-integrated flow fluctuations $\la v_n^4\ra/\la v_n^2\ra$, it does not depend on the magnitude of the flow coefficient itself. The flow magnitude fluctuations are strongest in the 0-5\% central collisions and then decrease as the centrality increases. The largest deviation from unity is observed for the Par3 calculation, which also deviates from unity at lower values of $\pt$ compared to the other calculations. The large deviation from unity suggests a strong sensitivity of the flow magnitude fluctuations to the initial conditions by changing the Lund string parameters with an effect of up to 20\% when comparing Par3 vs Par1. The flow magnitude fluctuations $M_2^\mathrm{f}$ show no sensitivity to the partonic cross section within uncertainties, regardless of whether ART is enabled (Par4 vs Par1) or not (Par5 vs Par2). The puzzling $\eta/s$-dependence shown in \cite{ALICE:2022smy} based on the models from \cite{Bozek:2021mov} is not observed in the low \pt region even though the change in $\eta/s$ is larger in the AMPT calculations. The AMPT calculations show \Fvn{2} = 1 at low \pt and the results are compatible for different $\sigma$, as expected. The hadronic rescattering plays a small role in the flow magnitude fluctuations with $\sim$10\% difference between the calculations with and without ART in the second-highest \pt bin, which has smaller uncertainties. Both the 1.5 mb (Par4 vs Par1) and 3.0 mb (Par5 vs Par2) calculations are equally affected by the hadronic phase within uncertainties. The Par1 AMPT model with $\sigma = 1.5$ mb, a = 0.3, b = 0.15 and ART ON describes the ALICE data reasonably well in all centralities. The calculation slightly underestimates the data at $\pt > 3.0$ GeV/\textit{c} in the 0-5\% and 5-10\% central collisions and at $\pt > 2.0$ GeV/\textit{c} in the 30-40\% and 40-50\% central collisions. The flow magnitude fluctuations are reproduced in the transport model, and the data vs model comparison enable a further opportunity to tune the initial conditions in AMPT and, in general, the use of ALICE data to constrain the initial conditions of the heavy-ion collisions. 
\section{\label{sec:level6}Summary}
In this paper, the AMPT calculations of \pt-dependent flow vector, flow angle and flow magnitude fluctuations in \PbPb collisions at \snn = 5.02 TeV are presented. The different configurations of the AMPT model probe the sensitivity of the fluctuations to 1) the initial conditions, 2) the partonic cross section, and 3) the hadronic rescattering to pinpoint the source of these fluctuations. It is found that the observables are highly sensitive to the changes in the initial conditions. The observables show weak to no dependence on changes in the partonic cross section and the hadronic rescattering. The high sensitivity to changes in the initial state suggests that the fluctuations mainly originate from the event-by-event fluctuations of the initial state and are only slightly dependent on the transport properties of the QGP. The presented studies on the \pt-dependent flow vector fluctuations and comparison between models and data can help constrain the initial stage of the QGP formation.
\begin{acknowledgments}
This work is supported by a research grant (00025462) from VILLUM FONDEN.
\end{acknowledgments}
\appendix
\section{\label{app:ratios}Model ratios}
\begin{figure*}[h!]
\centering
\includegraphics[scale=0.8]{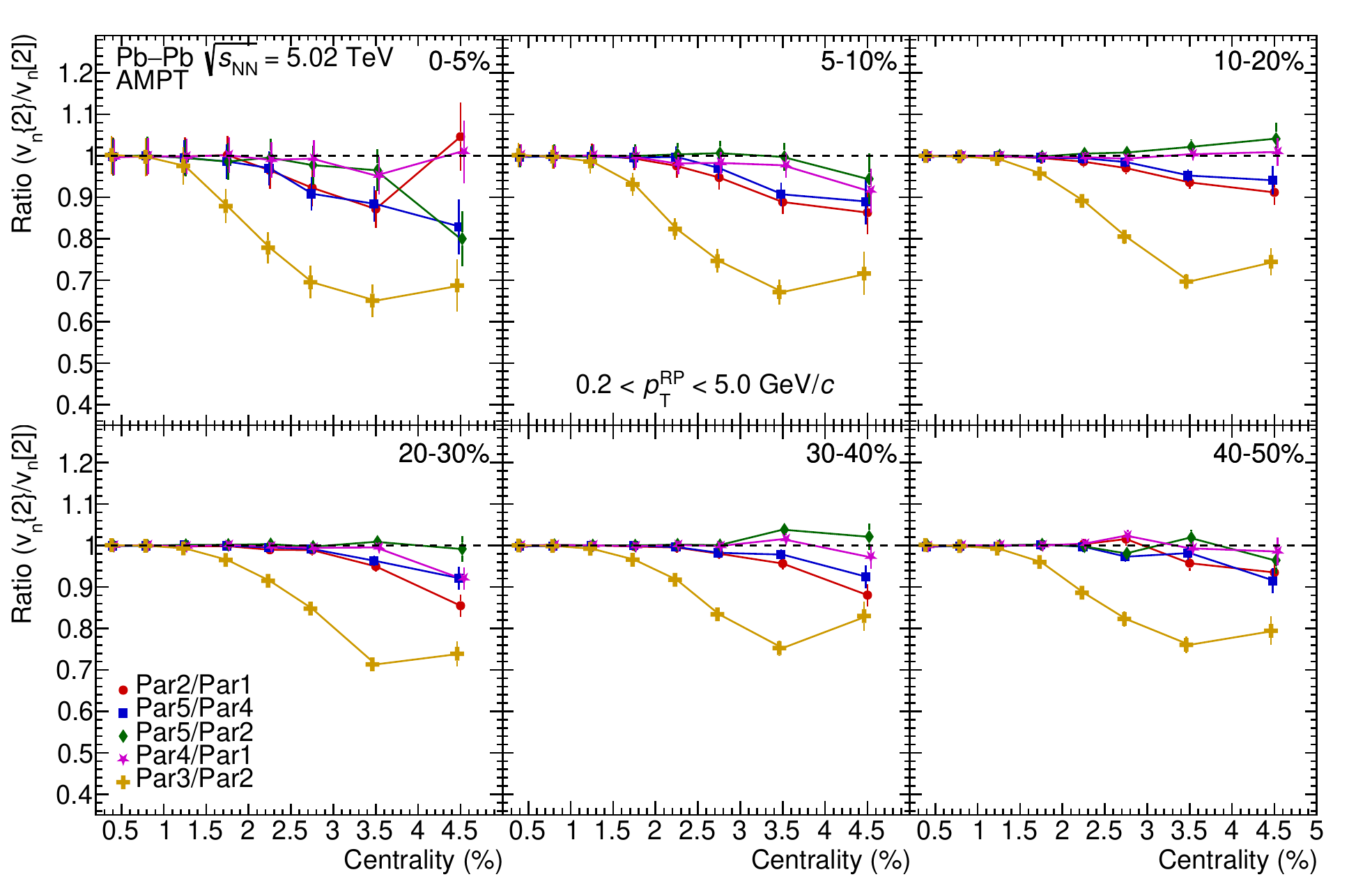}
\caption{\label{fig:v2ratio_ratio}The ratio of the AMPT calculations of $v_2\{2\}/v_2[2]$ with different configurations; ART OF vs ART ON with 1.5 mb (Par2/Par1), ART OF vs ART ON with 3.0 mb (Par5/Par3), 3.0 mb vs 1.5 mb with ART OFF (Par5/Par2), 3.0 mb vs 1.5 mb with ART ON (Par4/Par1), and $a=0.5, b=0.9$ vs $a=0.3, b=0.15$ with 1.5 mb and ART OFF (Par3/Par2).}
\end{figure*}

\begin{figure*}[h!]
\centering
\includegraphics[scale=0.8]{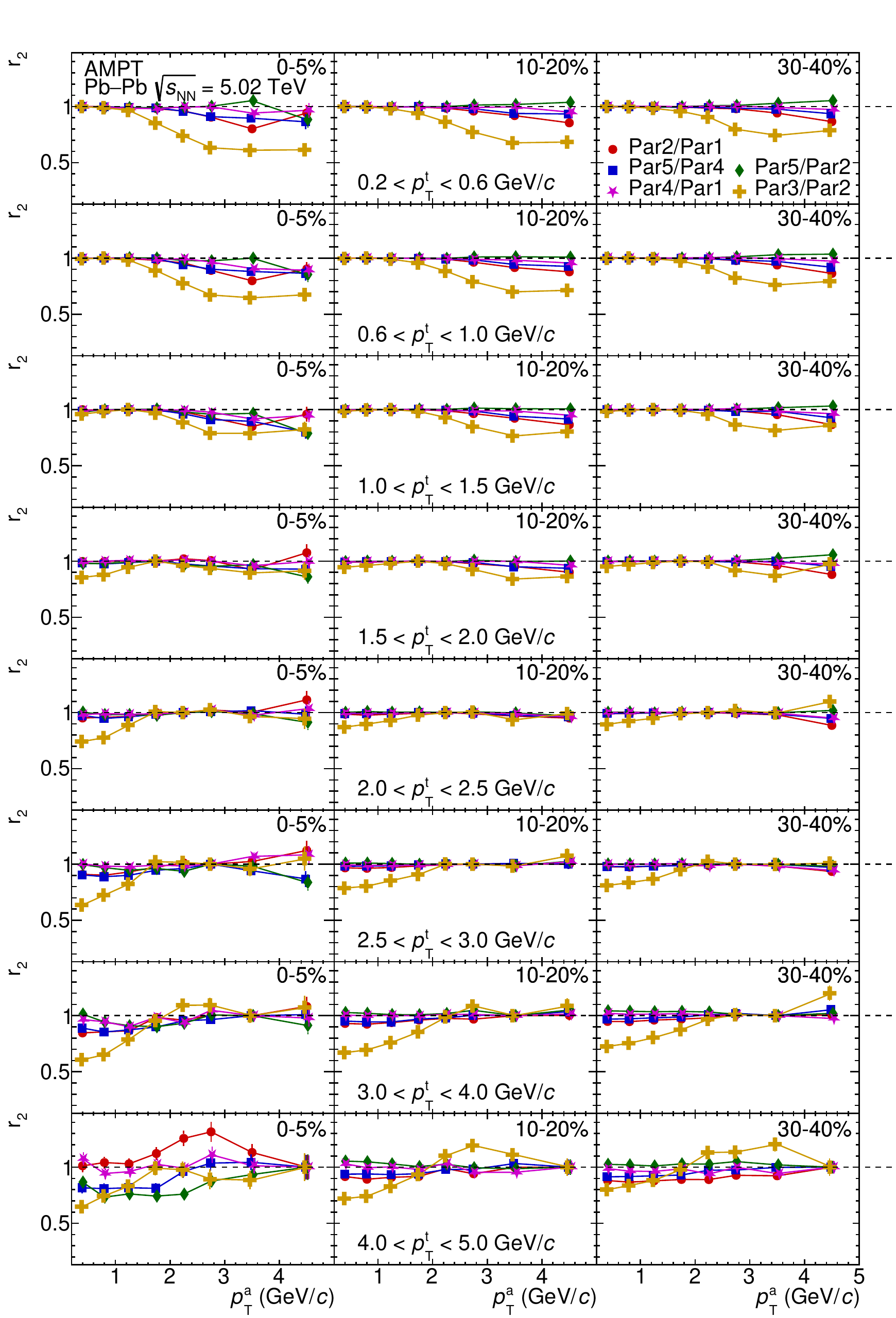}
\caption{\label{fig:r2_ratio}The ratio of the AMPT calculations of $r_2$ with different configurations; ART OF vs ART ON with 1.5 mb (Par2/Par1), ART OF vs ART ON with 3.0 mb (Par5/Par3), 3.0 mb vs 1.5 mb with ART OFF (Par5/Par2), 3.0 mb vs 1.5 mb with ART ON (Par4/Par1), and $a=0.5, b=0.9$ vs $a=0.3, b=0.15$ with 1.5 mb and ART OFF (Par3/Par2).}
\end{figure*}

\begin{figure*}[h!]
\centering
\includegraphics[scale=0.7]{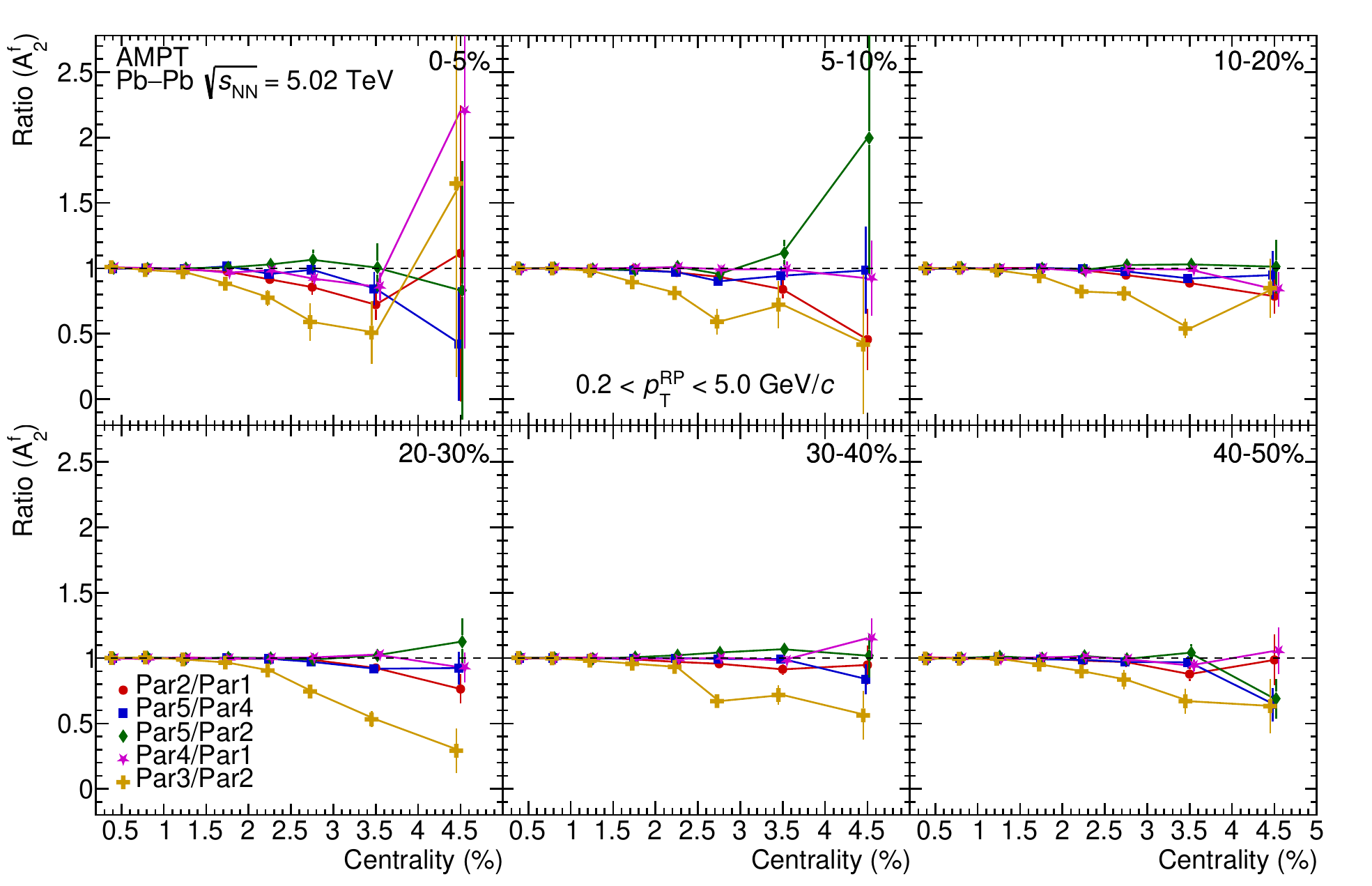}
\caption{\label{fig:A2f_ratio}The ratio of the AMPT calculations of $\Fpsi{2}$ with different configurations; ART OF vs ART ON with 1.5 mb (Par2/Par1), ART OF vs ART ON with 3.0 mb (Par5/Par3), 3.0 mb vs 1.5 mb with ART OFF (Par5/Par2), 3.0 mb vs 1.5 mb with ART ON (Par4/Par1), and $a=0.5, b=0.9$ vs $a=0.3, b=0.15$ with 1.5 mb and ART OFF (Par3/Par2).}
\end{figure*}

\begin{figure*}[h!]
\centering
\includegraphics[scale=0.7]{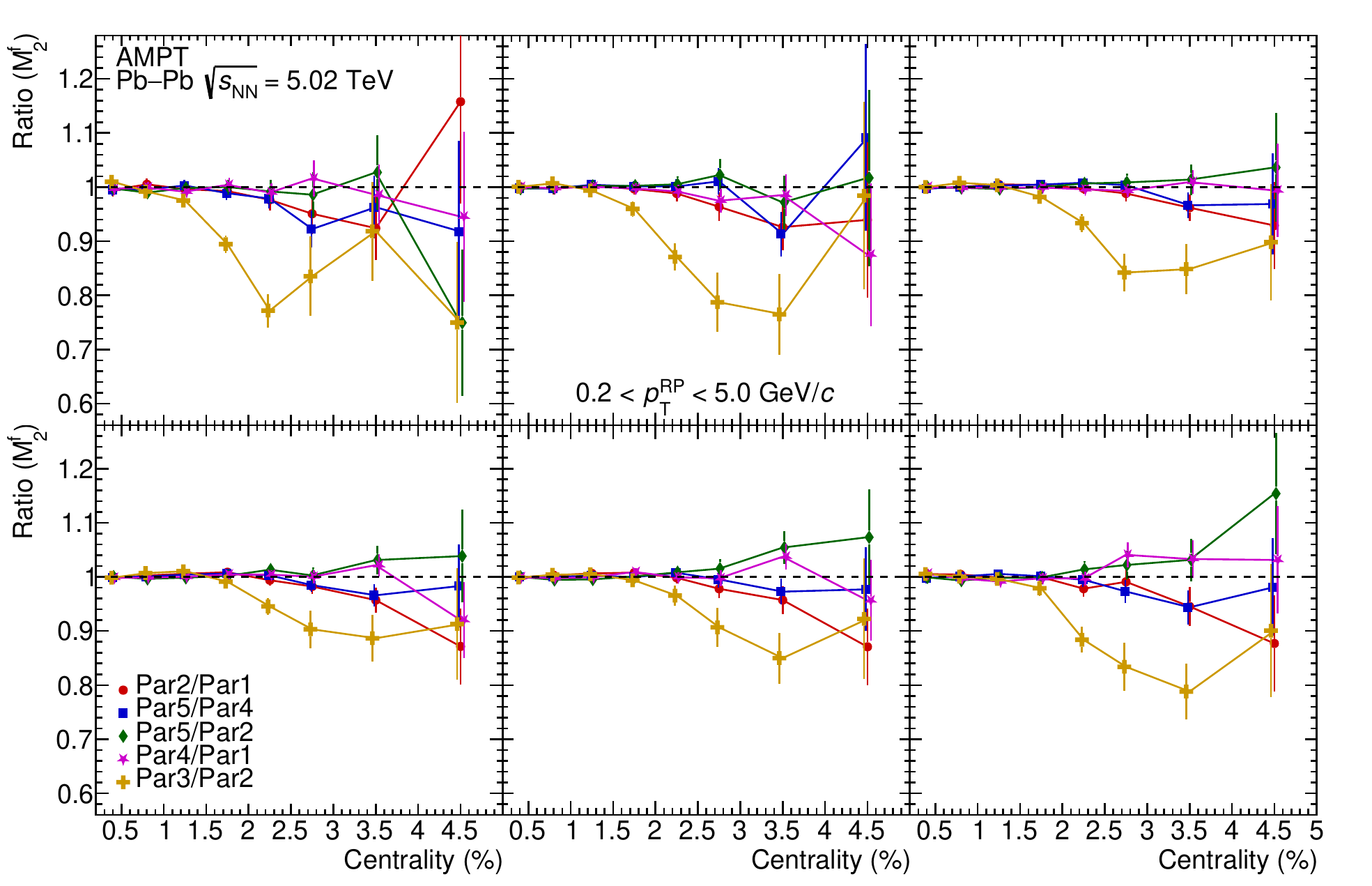}
\caption{\label{fig:M2f_ratio}The ratio of the AMPT calculations of $\Fvn{2}$ with different configurations; ART OF vs ART ON with 1.5 mb (Par2/Par1), ART OF vs ART ON with 3.0 mb (Par5/Par3), 3.0 mb vs 1.5 mb with ART OFF (Par5/Par2), 3.0 mb vs 1.5 mb with ART ON (Par4/Par1), and $a=0.5, b=0.9$ vs $a=0.3, b=0.15$ with 1.5 mb and ART OFF (Par3/Par2).}
\end{figure*}

In order to compare the different configurations against each other the ratios of the model calculations are presented here. Fig. \ref{fig:v2ratio_ratio} shows the comparison for $v_2\{2\}/v_2[2]$. The $r_2$ comparison is shown in fig. \ref{fig:r2_ratio}. Finally, the ratios for the flow angle fluctuations $\Fpsi{2}$ and the flow magnitude fluctuations $\Fvn{2}$ are shown in figures \ref{fig:A2f_ratio} and \ref{fig:M2f_ratio}, respectively.
\section{\label{app:rn}Factorization ratio}
\begin{figure*}
\centering
\includegraphics[scale=0.92]{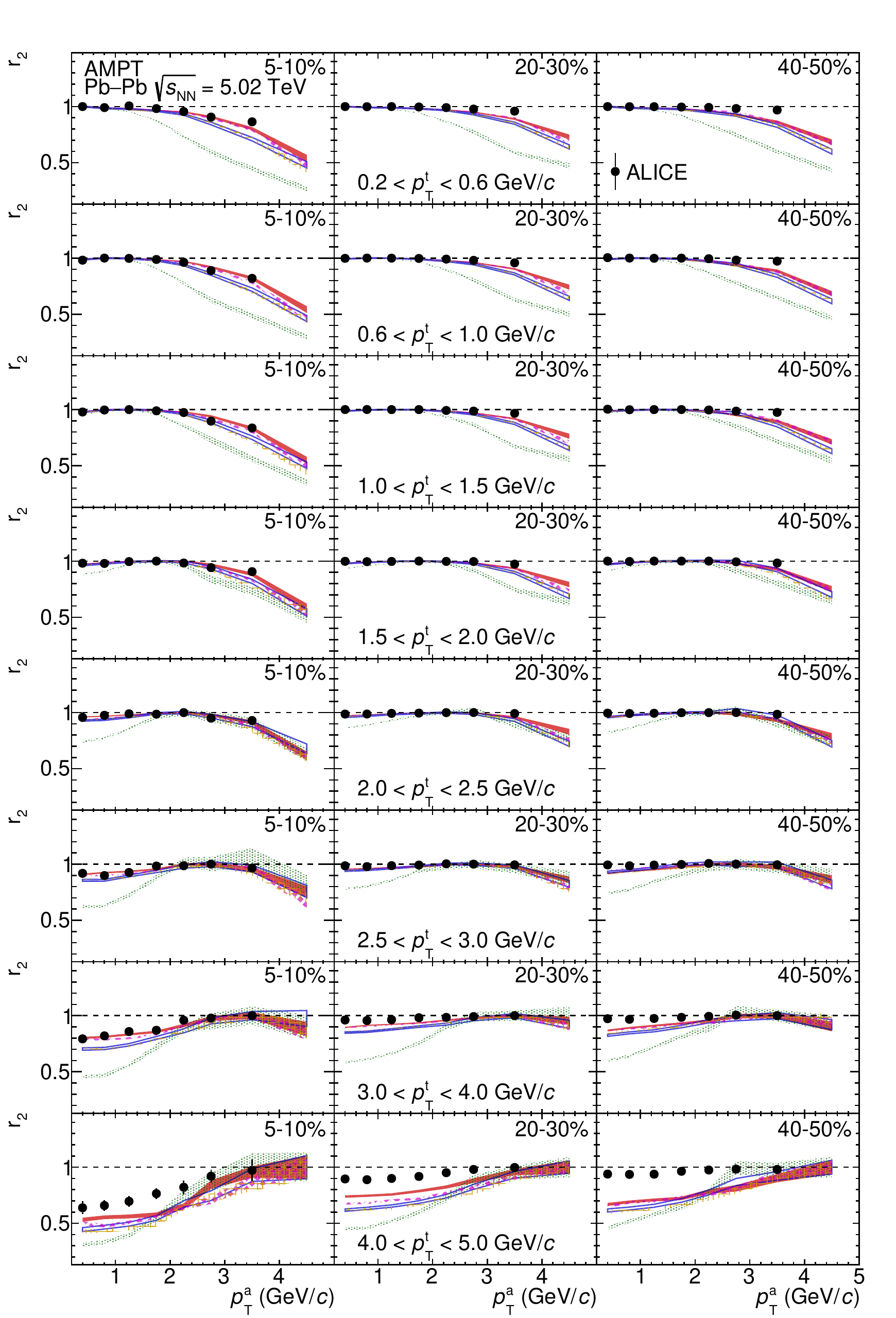}
\caption{\label{fig:r2_2}The \pt-dependence of the factorization ratio $r_2$ in centralities 5-10\%, 20-30\% and 40-50\% with different \textit{trigger} particle transverse momentum ranges from the AMPT model with different partonic cross sections, different Lund string parameters and with and without a hadronic rescattering phase. ALICE data points are shown as black circles for comparison.}
\end{figure*}

The factorization ratio of the second-order flow harmonic, $r_2$, is shown in fig. \ref{fig:r2_2} as a function of \textit{associate} particle transverse momentum, \pta, in centrality ranges 5-10\%, 20-30\% and 40-50\% and different \textit{trigger} particle transverse momentum, \ptt, ranges. These are the centrality ranges not presented in the main body of this paper.
\newpage
\bibliography{bibliography.bib}% Produces the bibliography via BibTeX.

%apsrev4-2.bst 2019-01-14 (MD) hand-edited version of apsrev4-1.bst
%Control: key (0)
%Control: author (8) initials jnrlst
%Control: editor formatted (1) identically to author
%Control: production of article title (0) allowed
%Control: page (0) single
%Control: year (1) truncated
%Control: production of eprint (0) enabled
\begin{thebibliography}{75}%
\makeatletter
\providecommand \@ifxundefined [1]{%
 \@ifx{#1\undefined}
}%
\providecommand \@ifnum [1]{%
 \ifnum #1\expandafter \@firstoftwo
 \else \expandafter \@secondoftwo
 \fi
}%
\providecommand \@ifx [1]{%
 \ifx #1\expandafter \@firstoftwo
 \else \expandafter \@secondoftwo
 \fi
}%
\providecommand \natexlab [1]{#1}%
\providecommand \enquote  [1]{``#1''}%
\providecommand \bibnamefont  [1]{#1}%
\providecommand \bibfnamefont [1]{#1}%
\providecommand \citenamefont [1]{#1}%
\providecommand \href@noop [0]{\@secondoftwo}%
\providecommand \href [0]{\begingroup \@sanitize@url \@href}%
\providecommand \@href[1]{\@@startlink{#1}\@@href}%
\providecommand \@@href[1]{\endgroup#1\@@endlink}%
\providecommand \@sanitize@url [0]{\catcode `\\12\catcode `\$12\catcode
  `\&12\catcode `\#12\catcode `\^12\catcode `\_12\catcode `\%12\relax}%
\providecommand \@@startlink[1]{}%
\providecommand \@@endlink[0]{}%
\providecommand \url  [0]{\begingroup\@sanitize@url \@url }%
\providecommand \@url [1]{\endgroup\@href {#1}{\urlprefix }}%
\providecommand \urlprefix  [0]{URL }%
\providecommand \Eprint [0]{\href }%
\providecommand \doibase [0]{https://doi.org/}%
\providecommand \selectlanguage [0]{\@gobble}%
\providecommand \bibinfo  [0]{\@secondoftwo}%
\providecommand \bibfield  [0]{\@secondoftwo}%
\providecommand \translation [1]{[#1]}%
\providecommand \BibitemOpen [0]{}%
\providecommand \bibitemStop [0]{}%
\providecommand \bibitemNoStop [0]{.\EOS\space}%
\providecommand \EOS [0]{\spacefactor3000\relax}%
\providecommand \BibitemShut  [1]{\csname bibitem#1\endcsname}%
\let\auto@bib@innerbib\@empty
%</preamble>
\bibitem [{\citenamefont {Arsene}\ \emph {et~al.}(2005)\citenamefont {Arsene}
  \emph {et~al.}}]{BRAHMS:2004adc}%
  \BibitemOpen
  \bibfield  {author} {\bibinfo {author} {\bibfnamefont {I.}~\bibnamefont
  {Arsene}} \emph {et~al.} (\bibinfo {collaboration} {BRAHMS}),\ }\bibfield
  {title} {\bibinfo {title} {{Quark gluon plasma and color glass condensate at
  RHIC? The Perspective from the BRAHMS experiment}},\ }\href
  {https://doi.org/10.1016/j.nuclphysa.2005.02.130} {\bibfield  {journal}
  {\bibinfo  {journal} {Nucl. Phys. A}\ }\textbf {\bibinfo {volume} {757}},\
  \bibinfo {pages} {1} (\bibinfo {year} {2005})},\ \Eprint
  {https://arxiv.org/abs/nucl-ex/0410020} {arXiv:nucl-ex/0410020} \BibitemShut
  {NoStop}%
\bibitem [{\citenamefont {Adams}\ \emph {et~al.}(2005)\citenamefont {Adams}
  \emph {et~al.}}]{STAR:2005gfr}%
  \BibitemOpen
  \bibfield  {author} {\bibinfo {author} {\bibfnamefont {J.}~\bibnamefont
  {Adams}} \emph {et~al.} (\bibinfo {collaboration} {STAR}),\ }\bibfield
  {title} {\bibinfo {title} {{Experimental and theoretical challenges in the
  search for the quark gluon plasma: The STAR Collaboration's critical
  assessment of the evidence from RHIC collisions}},\ }\href
  {https://doi.org/10.1016/j.nuclphysa.2005.03.085} {\bibfield  {journal}
  {\bibinfo  {journal} {Nucl. Phys. A}\ }\textbf {\bibinfo {volume} {757}},\
  \bibinfo {pages} {102} (\bibinfo {year} {2005})},\ \Eprint
  {https://arxiv.org/abs/nucl-ex/0501009} {arXiv:nucl-ex/0501009} \BibitemShut
  {NoStop}%
\bibitem [{\citenamefont {Adcox}\ \emph {et~al.}(2005)\citenamefont {Adcox}
  \emph {et~al.}}]{PHENIX:2004vcz}%
  \BibitemOpen
  \bibfield  {author} {\bibinfo {author} {\bibfnamefont {K.}~\bibnamefont
  {Adcox}} \emph {et~al.} (\bibinfo {collaboration} {PHENIX}),\ }\bibfield
  {title} {\bibinfo {title} {{Formation of dense partonic matter in
  relativistic nucleus-nucleus collisions at RHIC: Experimental evaluation by
  the PHENIX collaboration}},\ }\href
  {https://doi.org/10.1016/j.nuclphysa.2005.03.086} {\bibfield  {journal}
  {\bibinfo  {journal} {Nucl. Phys. A}\ }\textbf {\bibinfo {volume} {757}},\
  \bibinfo {pages} {184} (\bibinfo {year} {2005})},\ \Eprint
  {https://arxiv.org/abs/nucl-ex/0410003} {arXiv:nucl-ex/0410003} \BibitemShut
  {NoStop}%
\bibitem [{\citenamefont {Back}\ \emph {et~al.}(2005)\citenamefont {Back} \emph
  {et~al.}}]{PHOBOS:2004zne}%
  \BibitemOpen
  \bibfield  {author} {\bibinfo {author} {\bibfnamefont {B.~B.}\ \bibnamefont
  {Back}} \emph {et~al.} (\bibinfo {collaboration} {PHOBOS}),\ }\bibfield
  {title} {\bibinfo {title} {{The PHOBOS perspective on discoveries at RHIC}},\
  }\href {https://doi.org/10.1016/j.nuclphysa.2005.03.084} {\bibfield
  {journal} {\bibinfo  {journal} {Nucl. Phys. A}\ }\textbf {\bibinfo {volume}
  {757}},\ \bibinfo {pages} {28} (\bibinfo {year} {2005})},\ \Eprint
  {https://arxiv.org/abs/nucl-ex/0410022} {arXiv:nucl-ex/0410022} \BibitemShut
  {NoStop}%
\bibitem [{\citenamefont {Muller}\ \emph {et~al.}(2012)\citenamefont {Muller},
  \citenamefont {Schukraft},\ and\ \citenamefont {Wyslouch}}]{Muller:2012zq}%
  \BibitemOpen
  \bibfield  {author} {\bibinfo {author} {\bibfnamefont {B.}~\bibnamefont
  {Muller}}, \bibinfo {author} {\bibfnamefont {J.}~\bibnamefont {Schukraft}},\
  and\ \bibinfo {author} {\bibfnamefont {B.}~\bibnamefont {Wyslouch}},\
  }\bibfield  {title} {\bibinfo {title} {{First Results from Pb+Pb collisions
  at the LHC}},\ }\href {https://doi.org/10.1146/annurev-nucl-102711-094910}
  {\bibfield  {journal} {\bibinfo  {journal} {Ann. Rev. Nucl. Part. Sci.}\
  }\textbf {\bibinfo {volume} {62}},\ \bibinfo {pages} {361} (\bibinfo {year}
  {2012})},\ \Eprint {https://arxiv.org/abs/1202.3233} {arXiv:1202.3233
  [hep-ex]} \BibitemShut {NoStop}%
\bibitem [{\citenamefont {Ollitrault}(1992)}]{Ollitrault:1992bk}%
  \BibitemOpen
  \bibfield  {author} {\bibinfo {author} {\bibfnamefont {J.-Y.}\ \bibnamefont
  {Ollitrault}},\ }\bibfield  {title} {\bibinfo {title} {{Anisotropy as a
  signature of transverse collective flow}},\ }\href
  {https://doi.org/10.1103/PhysRevD.46.229} {\bibfield  {journal} {\bibinfo
  {journal} {Phys. Rev.}\ }\textbf {\bibinfo {volume} {D46}},\ \bibinfo {pages}
  {229} (\bibinfo {year} {1992})}\BibitemShut {NoStop}%
%%CITATION = PHRVA,D46,229;%%
\bibitem [{\citenamefont {Voloshin}\ \emph {et~al.}(2010)\citenamefont
  {Voloshin}, \citenamefont {Poskanzer},\ and\ \citenamefont
  {Snellings}}]{Voloshin:2008dg}%
  \BibitemOpen
  \bibfield  {author} {\bibinfo {author} {\bibfnamefont {S.~A.}\ \bibnamefont
  {Voloshin}}, \bibinfo {author} {\bibfnamefont {A.~M.}\ \bibnamefont
  {Poskanzer}},\ and\ \bibinfo {author} {\bibfnamefont {R.}~\bibnamefont
  {Snellings}},\ }\bibfield  {title} {\bibinfo {title} {{Collective phenomena
  in non-central nuclear collisions}},\ }\href
  {https://doi.org/10.1007/978-3-642-01539-7_10} {\bibfield  {journal}
  {\bibinfo  {journal} {Landolt-Bornstein}\ }\textbf {\bibinfo {volume} {23}},\
  \bibinfo {pages} {293} (\bibinfo {year} {2010})},\ \Eprint
  {https://arxiv.org/abs/0809.2949} {arXiv:0809.2949 [nucl-ex]} \BibitemShut
  {NoStop}%
\bibitem [{\citenamefont {Voloshin}\ and\ \citenamefont
  {Zhang}(1996)}]{Voloshin:1994mz}%
  \BibitemOpen
  \bibfield  {author} {\bibinfo {author} {\bibfnamefont {S.}~\bibnamefont
  {Voloshin}}\ and\ \bibinfo {author} {\bibfnamefont {Y.}~\bibnamefont
  {Zhang}},\ }\bibfield  {title} {\bibinfo {title} {{Flow study in relativistic
  nuclear collisions by Fourier expansion of Azimuthal particle
  distributions}},\ }\href {https://doi.org/10.1007/s002880050141} {\bibfield
  {journal} {\bibinfo  {journal} {Z. Phys.}\ }\textbf {\bibinfo {volume}
  {C70}},\ \bibinfo {pages} {665} (\bibinfo {year} {1996})},\ \Eprint
  {https://arxiv.org/abs/hep-ph/9407282} {arXiv:hep-ph/9407282 [hep-ph]}
  \BibitemShut {NoStop}%
%%CITATION = HEP-PH/9407282;%%
\bibitem [{\citenamefont {Ackermann}\ \emph {et~al.}(2001)\citenamefont
  {Ackermann} \emph {et~al.}}]{STAR:2000ekf}%
  \BibitemOpen
  \bibfield  {author} {\bibinfo {author} {\bibfnamefont {K.~H.}\ \bibnamefont
  {Ackermann}} \emph {et~al.} (\bibinfo {collaboration} {STAR}),\ }\bibfield
  {title} {\bibinfo {title} {{Elliptic flow in Au + Au collisions at
  (S(NN))**(1/2) = 130 GeV}},\ }\href
  {https://doi.org/10.1103/PhysRevLett.86.402} {\bibfield  {journal} {\bibinfo
  {journal} {Phys. Rev. Lett.}\ }\textbf {\bibinfo {volume} {86}},\ \bibinfo
  {pages} {402} (\bibinfo {year} {2001})},\ \Eprint
  {https://arxiv.org/abs/nucl-ex/0009011} {arXiv:nucl-ex/0009011} \BibitemShut
  {NoStop}%
\bibitem [{\citenamefont {Adler}\ \emph {et~al.}(2003)\citenamefont {Adler}
  \emph {et~al.}}]{PHENIX:2003qra}%
  \BibitemOpen
  \bibfield  {author} {\bibinfo {author} {\bibfnamefont {S.~S.}\ \bibnamefont
  {Adler}} \emph {et~al.} (\bibinfo {collaboration} {PHENIX}),\ }\bibfield
  {title} {\bibinfo {title} {{Elliptic flow of identified hadrons in Au+Au
  collisions at s(NN)**(1/2) = 200-GeV}},\ }\href
  {https://doi.org/10.1103/PhysRevLett.91.182301} {\bibfield  {journal}
  {\bibinfo  {journal} {Phys. Rev. Lett.}\ }\textbf {\bibinfo {volume} {91}},\
  \bibinfo {pages} {182301} (\bibinfo {year} {2003})},\ \Eprint
  {https://arxiv.org/abs/nucl-ex/0305013} {arXiv:nucl-ex/0305013} \BibitemShut
  {NoStop}%
\bibitem [{\citenamefont {Adamczyk}\ \emph {et~al.}(2013)\citenamefont
  {Adamczyk} \emph {et~al.}}]{STAR:2013qio}%
  \BibitemOpen
  \bibfield  {author} {\bibinfo {author} {\bibfnamefont {L.}~\bibnamefont
  {Adamczyk}} \emph {et~al.} (\bibinfo {collaboration} {STAR}),\ }\bibfield
  {title} {\bibinfo {title} {{Third Harmonic Flow of Charged Particles in Au+Au
  Collisions at sqrtsNN = 200 GeV}},\ }\href
  {https://doi.org/10.1103/PhysRevC.88.014904} {\bibfield  {journal} {\bibinfo
  {journal} {Phys. Rev. C}\ }\textbf {\bibinfo {volume} {88}},\ \bibinfo
  {pages} {014904} (\bibinfo {year} {2013})},\ \Eprint
  {https://arxiv.org/abs/1301.2187} {arXiv:1301.2187 [nucl-ex]} \BibitemShut
  {NoStop}%
\bibitem [{\citenamefont {Adare}\ \emph {et~al.}(2015)\citenamefont {Adare}
  \emph {et~al.}}]{PHENIX:2015idk}%
  \BibitemOpen
  \bibfield  {author} {\bibinfo {author} {\bibfnamefont {A.}~\bibnamefont
  {Adare}} \emph {et~al.} (\bibinfo {collaboration} {PHENIX}),\ }\bibfield
  {title} {\bibinfo {title} {{Measurements of elliptic and triangular flow in
  high-multiplicity $^{3}$He$+$Au collisions at $\sqrt{s_{_{NN}}}=200$ GeV}},\
  }\href {https://doi.org/10.1103/PhysRevLett.115.142301} {\bibfield  {journal}
  {\bibinfo  {journal} {Phys. Rev. Lett.}\ }\textbf {\bibinfo {volume} {115}},\
  \bibinfo {pages} {142301} (\bibinfo {year} {2015})},\ \Eprint
  {https://arxiv.org/abs/1507.06273} {arXiv:1507.06273 [nucl-ex]} \BibitemShut
  {NoStop}%
\bibitem [{\citenamefont {Aamodt}\ \emph {et~al.}(2010)\citenamefont {Aamodt}
  \emph {et~al.}}]{Aamodt:2010pa}%
  \BibitemOpen
  \bibfield  {author} {\bibinfo {author} {\bibfnamefont {K.}~\bibnamefont
  {Aamodt}} \emph {et~al.} (\bibinfo {collaboration} {ALICE}),\ }\bibfield
  {title} {\bibinfo {title} {{Elliptic flow of charged particles in Pb-Pb
  collisions at 2.76 TeV}},\ }\href
  {https://doi.org/10.1103/PhysRevLett.105.252302} {\bibfield  {journal}
  {\bibinfo  {journal} {Phys. Rev. Lett.}\ }\textbf {\bibinfo {volume} {105}},\
  \bibinfo {pages} {252302} (\bibinfo {year} {2010})},\ \Eprint
  {https://arxiv.org/abs/1011.3914} {arXiv:1011.3914 [nucl-ex]} \BibitemShut
  {NoStop}%
%%CITATION = ARXIV:1011.3914;%%
\bibitem [{\citenamefont {Aamodt}\ \emph {et~al.}(2011)\citenamefont {Aamodt}
  \emph {et~al.}}]{ALICE:2011ab}%
  \BibitemOpen
  \bibfield  {author} {\bibinfo {author} {\bibfnamefont {K.}~\bibnamefont
  {Aamodt}} \emph {et~al.} (\bibinfo {collaboration} {ALICE}),\ }\bibfield
  {title} {\bibinfo {title} {{Higher harmonic anisotropic flow measurements of
  charged particles in Pb-Pb collisions at $\sqrt{s_{NN}}$=2.76 TeV}},\ }\href
  {https://doi.org/10.1103/PhysRevLett.107.032301} {\bibfield  {journal}
  {\bibinfo  {journal} {Phys. Rev. Lett.}\ }\textbf {\bibinfo {volume} {107}},\
  \bibinfo {pages} {032301} (\bibinfo {year} {2011})},\ \Eprint
  {https://arxiv.org/abs/1105.3865} {arXiv:1105.3865 [nucl-ex]} \BibitemShut
  {NoStop}%
%%CITATION = ARXIV:1105.3865;%%
\bibitem [{\citenamefont {Abelev}\ \emph {et~al.}(2015)\citenamefont {Abelev}
  \emph {et~al.}}]{Abelev:2014pua}%
  \BibitemOpen
  \bibfield  {author} {\bibinfo {author} {\bibfnamefont {B.}~\bibnamefont
  {Abelev}} \emph {et~al.} (\bibinfo {collaboration} {ALICE}),\ }\bibfield
  {title} {\bibinfo {title} {{Elliptic flow of identified hadrons in Pb-Pb
  collisions at $ \sqrt{s_{\mathrm{NN}}}=2.76 $ TeV}},\ }\href
  {https://doi.org/10.1007/JHEP06(2015)190} {\bibfield  {journal} {\bibinfo
  {journal} {JHEP}\ }\textbf {\bibinfo {volume} {06}},\ \bibinfo {pages}
  {190}},\ \Eprint {https://arxiv.org/abs/1405.4632} {arXiv:1405.4632
  [nucl-ex]} \BibitemShut {NoStop}%
%%CITATION = ARXIV:1405.4632;%%
\bibitem [{\citenamefont {Adam}\ \emph {et~al.}(2016)\citenamefont {Adam} \emph
  {et~al.}}]{Adam:2016izf}%
  \BibitemOpen
  \bibfield  {author} {\bibinfo {author} {\bibfnamefont {J.}~\bibnamefont
  {Adam}} \emph {et~al.} (\bibinfo {collaboration} {ALICE}),\ }\bibfield
  {title} {\bibinfo {title} {{Anisotropic flow of charged particles in Pb-Pb
  collisions at $\sqrt{s_{\rm NN}}=5.02$ TeV}},\ }\href
  {https://doi.org/10.1103/PhysRevLett.116.132302} {\bibfield  {journal}
  {\bibinfo  {journal} {Phys. Rev. Lett.}\ }\textbf {\bibinfo {volume} {116}},\
  \bibinfo {pages} {132302} (\bibinfo {year} {2016})},\ \Eprint
  {https://arxiv.org/abs/1602.01119} {arXiv:1602.01119 [nucl-ex]} \BibitemShut
  {NoStop}%
%%CITATION = ARXIV:1602.01119;%%
\bibitem [{\citenamefont {Acharya}\ \emph
  {et~al.}(2017{\natexlab{a}})\citenamefont {Acharya} \emph
  {et~al.}}]{Acharya:2017zfg}%
  \BibitemOpen
  \bibfield  {author} {\bibinfo {author} {\bibfnamefont {S.}~\bibnamefont
  {Acharya}} \emph {et~al.} (\bibinfo {collaboration} {ALICE}),\ }\bibfield
  {title} {\bibinfo {title} {{Linear and non-linear flow modes in Pb-Pb
  collisions at $\sqrt{s_{\rm NN}} =$ 2.76 TeV}},\ }\href
  {https://doi.org/10.1016/j.physletb.2017.07.060} {\bibfield  {journal}
  {\bibinfo  {journal} {Phys. Lett.}\ }\textbf {\bibinfo {volume} {B773}},\
  \bibinfo {pages} {68} (\bibinfo {year} {2017}{\natexlab{a}})},\ \Eprint
  {https://arxiv.org/abs/1705.04377} {arXiv:1705.04377 [nucl-ex]} \BibitemShut
  {NoStop}%
%%CITATION = ARXIV:1705.04377;%%
\bibitem [{\citenamefont {Aad}\ \emph {et~al.}(2012{\natexlab{a}})\citenamefont
  {Aad} \emph {et~al.}}]{ATLAS:2012at}%
  \BibitemOpen
  \bibfield  {author} {\bibinfo {author} {\bibfnamefont {G.}~\bibnamefont
  {Aad}} \emph {et~al.} (\bibinfo {collaboration} {ATLAS}),\ }\bibfield
  {title} {\bibinfo {title} {{Measurement of the azimuthal anisotropy for
  charged particle production in $\sqrt{s_{NN}}=2.76$ TeV lead-lead collisions
  with the ATLAS detector}},\ }\href
  {https://doi.org/10.1103/PhysRevC.86.014907} {\bibfield  {journal} {\bibinfo
  {journal} {Phys. Rev.}\ }\textbf {\bibinfo {volume} {C86}},\ \bibinfo {pages}
  {014907} (\bibinfo {year} {2012}{\natexlab{a}})},\ \Eprint
  {https://arxiv.org/abs/1203.3087} {arXiv:1203.3087 [hep-ex]} \BibitemShut
  {NoStop}%
%%CITATION = ARXIV:1203.3087;%%
\bibitem [{\citenamefont {Aad}\ \emph {et~al.}(2012{\natexlab{b}})\citenamefont
  {Aad} \emph {et~al.}}]{ATLAS:2011ah}%
  \BibitemOpen
  \bibfield  {author} {\bibinfo {author} {\bibfnamefont {G.}~\bibnamefont
  {Aad}} \emph {et~al.} (\bibinfo {collaboration} {ATLAS}),\ }\bibfield
  {title} {\bibinfo {title} {{Measurement of the pseudorapidity and transverse
  momentum dependence of the elliptic flow of charged particles in lead-lead
  collisions at $\sqrt{s_{NN}}=2.76$ TeV with the ATLAS detector}},\ }\href
  {https://doi.org/10.1016/j.physletb.2011.12.056} {\bibfield  {journal}
  {\bibinfo  {journal} {Phys. Lett.}\ }\textbf {\bibinfo {volume} {B707}},\
  \bibinfo {pages} {330} (\bibinfo {year} {2012}{\natexlab{b}})},\ \Eprint
  {https://arxiv.org/abs/1108.6018} {arXiv:1108.6018 [hep-ex]} \BibitemShut
  {NoStop}%
%%CITATION = ARXIV:1108.6018;%%
\bibitem [{\citenamefont {Aad}\ \emph {et~al.}(2013)\citenamefont {Aad} \emph
  {et~al.}}]{Aad:2013xma}%
  \BibitemOpen
  \bibfield  {author} {\bibinfo {author} {\bibfnamefont {G.}~\bibnamefont
  {Aad}} \emph {et~al.} (\bibinfo {collaboration} {ATLAS}),\ }\bibfield
  {title} {\bibinfo {title} {{Measurement of the distributions of
  event-by-event flow harmonics in lead-lead collisions at
  $\sqrt{s_\mathrm{NN}} = 2.76$ TeV with the ATLAS detector at the LHC}},\
  }\href {https://doi.org/10.1007/JHEP11(2013)183} {\bibfield  {journal}
  {\bibinfo  {journal} {JHEP}\ }\textbf {\bibinfo {volume} {11}},\ \bibinfo
  {pages} {183}},\ \Eprint {https://arxiv.org/abs/1305.2942} {arXiv:1305.2942
  [hep-ex]} \BibitemShut {NoStop}%
\bibitem [{\citenamefont {Chatrchyan}\ \emph
  {et~al.}(2012{\natexlab{a}})\citenamefont {Chatrchyan} \emph
  {et~al.}}]{Chatrchyan:2012wg}%
  \BibitemOpen
  \bibfield  {author} {\bibinfo {author} {\bibfnamefont {S.}~\bibnamefont
  {Chatrchyan}} \emph {et~al.} (\bibinfo {collaboration} {CMS}),\ }\bibfield
  {title} {\bibinfo {title} {{Centrality dependence of dihadron correlations
  and azimuthal anisotropy harmonics in PbPb collisions at $\sqrt{s_{NN}}=2.76$
  TeV}},\ }\href {https://doi.org/10.1140/epjc/s10052-012-2012-3} {\bibfield
  {journal} {\bibinfo  {journal} {Eur. Phys. J.}\ }\textbf {\bibinfo {volume}
  {C72}},\ \bibinfo {pages} {2012} (\bibinfo {year} {2012}{\natexlab{a}})},\
  \Eprint {https://arxiv.org/abs/1201.3158} {arXiv:1201.3158 [nucl-ex]}
  \BibitemShut {NoStop}%
%%CITATION = ARXIV:1201.3158;%%
\bibitem [{\citenamefont {Chatrchyan}\ \emph {et~al.}(2013)\citenamefont
  {Chatrchyan} \emph {et~al.}}]{CMS:2012zex}%
  \BibitemOpen
  \bibfield  {author} {\bibinfo {author} {\bibfnamefont {S.}~\bibnamefont
  {Chatrchyan}} \emph {et~al.} (\bibinfo {collaboration} {CMS}),\ }\bibfield
  {title} {\bibinfo {title} {{Measurement of the elliptic anisotropy of charged
  particles produced in PbPb collisions at $\sqrt{s}_{NN}$=2.76 TeV}},\ }\href
  {https://doi.org/10.1103/PhysRevC.87.014902} {\bibfield  {journal} {\bibinfo
  {journal} {Phys. Rev.}\ }\textbf {\bibinfo {volume} {C87}},\ \bibinfo {pages}
  {014902} (\bibinfo {year} {2013})},\ \Eprint
  {https://arxiv.org/abs/1204.1409} {arXiv:1204.1409 [nucl-ex]} \BibitemShut
  {NoStop}%
\bibitem [{\citenamefont {Chatrchyan}\ \emph
  {et~al.}(2012{\natexlab{b}})\citenamefont {Chatrchyan} \emph
  {et~al.}}]{Chatrchyan:2012xq}%
  \BibitemOpen
  \bibfield  {author} {\bibinfo {author} {\bibfnamefont {S.}~\bibnamefont
  {Chatrchyan}} \emph {et~al.} (\bibinfo {collaboration} {CMS}),\ }\bibfield
  {title} {\bibinfo {title} {{Azimuthal anisotropy of charged particles at high
  transverse momenta in PbPb collisions at $\sqrt{s_{NN}}=2.76$ TeV}},\ }\href
  {https://doi.org/10.1103/PhysRevLett.109.022301} {\bibfield  {journal}
  {\bibinfo  {journal} {Phys. Rev. Lett.}\ }\textbf {\bibinfo {volume} {109}},\
  \bibinfo {pages} {022301} (\bibinfo {year} {2012}{\natexlab{b}})},\ \Eprint
  {https://arxiv.org/abs/1204.1850} {arXiv:1204.1850 [nucl-ex]} \BibitemShut
  {NoStop}%
\bibitem [{\citenamefont {Heinz}\ and\ \citenamefont
  {Snellings}(2013)}]{Heinz:2013th}%
  \BibitemOpen
  \bibfield  {author} {\bibinfo {author} {\bibfnamefont {U.}~\bibnamefont
  {Heinz}}\ and\ \bibinfo {author} {\bibfnamefont {R.}~\bibnamefont
  {Snellings}},\ }\bibfield  {title} {\bibinfo {title} {{Collective flow and
  viscosity in relativistic heavy-ion collisions}},\ }\href
  {https://doi.org/10.1146/annurev-nucl-102212-170540} {\bibfield  {journal}
  {\bibinfo  {journal} {Ann. Rev. Nucl. Part. Sci.}\ }\textbf {\bibinfo
  {volume} {63}},\ \bibinfo {pages} {123} (\bibinfo {year} {2013})},\ \Eprint
  {https://arxiv.org/abs/1301.2826} {arXiv:1301.2826 [nucl-th]} \BibitemShut
  {NoStop}%
%%CITATION = ARXIV:1301.2826;%%
\bibitem [{\citenamefont {Luzum}\ and\ \citenamefont
  {Petersen}(2014)}]{Luzum:2013yya}%
  \BibitemOpen
  \bibfield  {author} {\bibinfo {author} {\bibfnamefont {M.}~\bibnamefont
  {Luzum}}\ and\ \bibinfo {author} {\bibfnamefont {H.}~\bibnamefont
  {Petersen}},\ }\bibfield  {title} {\bibinfo {title} {{Initial State
  Fluctuations and Final State Correlations in Relativistic Heavy-Ion
  Collisions}},\ }\href {https://doi.org/10.1088/0954-3899/41/6/063102}
  {\bibfield  {journal} {\bibinfo  {journal} {J. Phys.}\ }\textbf {\bibinfo
  {volume} {G41}},\ \bibinfo {pages} {063102} (\bibinfo {year} {2014})},\
  \Eprint {https://arxiv.org/abs/1312.5503} {arXiv:1312.5503 [nucl-th]}
  \BibitemShut {NoStop}%
%%CITATION = ARXIV:1312.5503;%%
\bibitem [{\citenamefont {Shuryak}(2017)}]{Shuryak:2014zxa}%
  \BibitemOpen
  \bibfield  {author} {\bibinfo {author} {\bibfnamefont {E.}~\bibnamefont
  {Shuryak}},\ }\bibfield  {title} {\bibinfo {title} {{Strongly coupled
  quark-gluon plasma in heavy ion collisions}},\ }\href
  {https://doi.org/10.1103/RevModPhys.89.035001} {\bibfield  {journal}
  {\bibinfo  {journal} {Rev. Mod. Phys.}\ }\textbf {\bibinfo {volume} {89}},\
  \bibinfo {pages} {035001} (\bibinfo {year} {2017})},\ \Eprint
  {https://arxiv.org/abs/1412.8393} {arXiv:1412.8393 [hep-ph]} \BibitemShut
  {NoStop}%
%%CITATION = ARXIV:1412.8393;%%
\bibitem [{\citenamefont {Song}\ \emph {et~al.}(2017)\citenamefont {Song},
  \citenamefont {Zhou},\ and\ \citenamefont {Gajdosova}}]{Song:2017wtw}%
  \BibitemOpen
  \bibfield  {author} {\bibinfo {author} {\bibfnamefont {H.}~\bibnamefont
  {Song}}, \bibinfo {author} {\bibfnamefont {Y.}~\bibnamefont {Zhou}},\ and\
  \bibinfo {author} {\bibfnamefont {K.}~\bibnamefont {Gajdosova}},\ }\bibfield
  {title} {\bibinfo {title} {{Collective flow and hydrodynamics in large and
  small systems at the LHC}},\ }\href
  {https://doi.org/10.1007/s41365-017-0245-4} {\bibfield  {journal} {\bibinfo
  {journal} {Nucl. Sci. Tech.}\ }\textbf {\bibinfo {volume} {28}},\ \bibinfo
  {pages} {99} (\bibinfo {year} {2017})},\ \Eprint
  {https://arxiv.org/abs/1703.00670} {arXiv:1703.00670 [nucl-th]} \BibitemShut
  {NoStop}%
%%CITATION = ARXIV:1703.00670;%%
\bibitem [{\citenamefont {Bernhard}\ \emph {et~al.}(2019)\citenamefont
  {Bernhard}, \citenamefont {Moreland},\ and\ \citenamefont
  {Bass}}]{Bernhard:2019bmu}%
  \BibitemOpen
  \bibfield  {author} {\bibinfo {author} {\bibfnamefont {J.~E.}\ \bibnamefont
  {Bernhard}}, \bibinfo {author} {\bibfnamefont {J.~S.}\ \bibnamefont
  {Moreland}},\ and\ \bibinfo {author} {\bibfnamefont {S.~A.}\ \bibnamefont
  {Bass}},\ }\bibfield  {title} {\bibinfo {title} {{Bayesian estimation of the
  specific shear and bulk viscosity of quark\textendash{}gluon plasma}},\
  }\href {https://doi.org/10.1038/s41567-019-0611-8} {\bibfield  {journal}
  {\bibinfo  {journal} {Nature Phys.}\ }\textbf {\bibinfo {volume} {15}},\
  \bibinfo {pages} {1113} (\bibinfo {year} {2019})}\BibitemShut {NoStop}%
\bibitem [{\citenamefont {Everett}\ \emph {et~al.}(2021)\citenamefont {Everett}
  \emph {et~al.}}]{Everett:2020xug}%
  \BibitemOpen
  \bibfield  {author} {\bibinfo {author} {\bibfnamefont {D.}~\bibnamefont
  {Everett}} \emph {et~al.} (\bibinfo {collaboration} {JETSCAPE}),\ }\bibfield
  {title} {\bibinfo {title} {{Multisystem Bayesian constraints on the transport
  coefficients of QCD matter}},\ }\href
  {https://doi.org/10.1103/PhysRevC.103.054904} {\bibfield  {journal} {\bibinfo
   {journal} {Phys. Rev. C}\ }\textbf {\bibinfo {volume} {103}},\ \bibinfo
  {pages} {054904} (\bibinfo {year} {2021})},\ \Eprint
  {https://arxiv.org/abs/2011.01430} {arXiv:2011.01430 [hep-ph]} \BibitemShut
  {NoStop}%
\bibitem [{\citenamefont {Nijs}\ \emph {et~al.}(2021)\citenamefont {Nijs},
  \citenamefont {van~der Schee}, \citenamefont {G\"ursoy},\ and\ \citenamefont
  {Snellings}}]{Nijs:2020roc}%
  \BibitemOpen
  \bibfield  {author} {\bibinfo {author} {\bibfnamefont {G.}~\bibnamefont
  {Nijs}}, \bibinfo {author} {\bibfnamefont {W.}~\bibnamefont {van~der Schee}},
  \bibinfo {author} {\bibfnamefont {U.}~\bibnamefont {G\"ursoy}},\ and\
  \bibinfo {author} {\bibfnamefont {R.}~\bibnamefont {Snellings}},\ }\bibfield
  {title} {\bibinfo {title} {{Bayesian analysis of heavy ion collisions with
  the heavy ion computational framework Trajectum}},\ }\href
  {https://doi.org/10.1103/PhysRevC.103.054909} {\bibfield  {journal} {\bibinfo
   {journal} {Phys. Rev. C}\ }\textbf {\bibinfo {volume} {103}},\ \bibinfo
  {pages} {054909} (\bibinfo {year} {2021})},\ \Eprint
  {https://arxiv.org/abs/2010.15134} {arXiv:2010.15134 [nucl-th]} \BibitemShut
  {NoStop}%
\bibitem [{\citenamefont {Parkkila}\ \emph {et~al.}(2021)\citenamefont
  {Parkkila}, \citenamefont {Onnerstad}, \citenamefont {Taghavi}, \citenamefont
  {Mordasini}, \citenamefont {Bilandzic},\ and\ \citenamefont
  {Kim}}]{Parkkila:2021yha}%
  \BibitemOpen
  \bibfield  {author} {\bibinfo {author} {\bibfnamefont {J.~E.}\ \bibnamefont
  {Parkkila}}, \bibinfo {author} {\bibfnamefont {A.}~\bibnamefont {Onnerstad}},
  \bibinfo {author} {\bibfnamefont {S.~F.}\ \bibnamefont {Taghavi}}, \bibinfo
  {author} {\bibfnamefont {C.}~\bibnamefont {Mordasini}}, \bibinfo {author}
  {\bibfnamefont {A.}~\bibnamefont {Bilandzic}},\ and\ \bibinfo {author}
  {\bibfnamefont {D.~J.}\ \bibnamefont {Kim}},\ }\bibfield  {title} {\bibinfo
  {title} {{New constraints for QCD matter from improved Bayesian parameter
  estimation in heavy-ion collisions at LHC}},\ }\href@noop {} {\  (\bibinfo
  {year} {2021})},\ \Eprint {https://arxiv.org/abs/2111.08145}
  {arXiv:2111.08145 [hep-ph]} \BibitemShut {NoStop}%
\bibitem [{\citenamefont {Heinz}\ \emph {et~al.}(2013)\citenamefont {Heinz},
  \citenamefont {Qiu},\ and\ \citenamefont {Shen}}]{Heinz:2013bua}%
  \BibitemOpen
  \bibfield  {author} {\bibinfo {author} {\bibfnamefont {U.}~\bibnamefont
  {Heinz}}, \bibinfo {author} {\bibfnamefont {Z.}~\bibnamefont {Qiu}},\ and\
  \bibinfo {author} {\bibfnamefont {C.}~\bibnamefont {Shen}},\ }\bibfield
  {title} {\bibinfo {title} {{Fluctuating flow angles and anisotropic flow
  measurements}},\ }\href {https://doi.org/10.1103/PhysRevC.87.034913}
  {\bibfield  {journal} {\bibinfo  {journal} {Phys. Rev.}\ }\textbf {\bibinfo
  {volume} {C87}},\ \bibinfo {pages} {034913} (\bibinfo {year} {2013})},\
  \Eprint {https://arxiv.org/abs/1302.3535} {arXiv:1302.3535 [nucl-th]}
  \BibitemShut {NoStop}%
%%CITATION = ARXIV:1302.3535;%%
\bibitem [{\citenamefont {Gardim}\ \emph {et~al.}(2013)\citenamefont {Gardim},
  \citenamefont {Grassi}, \citenamefont {Luzum},\ and\ \citenamefont
  {Ollitrault}}]{Gardim:2012im}%
  \BibitemOpen
  \bibfield  {author} {\bibinfo {author} {\bibfnamefont {F.~G.}\ \bibnamefont
  {Gardim}}, \bibinfo {author} {\bibfnamefont {F.}~\bibnamefont {Grassi}},
  \bibinfo {author} {\bibfnamefont {M.}~\bibnamefont {Luzum}},\ and\ \bibinfo
  {author} {\bibfnamefont {J.-Y.}\ \bibnamefont {Ollitrault}},\ }\bibfield
  {title} {\bibinfo {title} {{Breaking of factorization of two-particle
  correlations in hydrodynamics}},\ }\href
  {https://doi.org/10.1103/PhysRevC.87.031901} {\bibfield  {journal} {\bibinfo
  {journal} {Phys. Rev.}\ }\textbf {\bibinfo {volume} {C87}},\ \bibinfo {pages}
  {031901} (\bibinfo {year} {2013})},\ \Eprint
  {https://arxiv.org/abs/1211.0989} {arXiv:1211.0989 [nucl-th]} \BibitemShut
  {NoStop}%
%%CITATION = ARXIV:1211.0989;%%
\bibitem [{\citenamefont {Acharya}\ \emph
  {et~al.}(2017{\natexlab{b}})\citenamefont {Acharya} \emph
  {et~al.}}]{ALICE:2017lyf}%
  \BibitemOpen
  \bibfield  {author} {\bibinfo {author} {\bibfnamefont {S.}~\bibnamefont
  {Acharya}} \emph {et~al.} (\bibinfo {collaboration} {ALICE}),\ }\bibfield
  {title} {\bibinfo {title} {{Searches for transverse momentum dependent flow
  vector fluctuations in Pb-Pb and p-Pb collisions at the LHC}},\ }\href
  {https://doi.org/10.1007/JHEP09(2017)032} {\bibfield  {journal} {\bibinfo
  {journal} {JHEP}\ }\textbf {\bibinfo {volume} {09}},\ \bibinfo {pages}
  {032}},\ \Eprint {https://arxiv.org/abs/1707.05690} {arXiv:1707.05690
  [nucl-ex]} \BibitemShut {NoStop}%
\bibitem [{\citenamefont {Chatrchyan}\ \emph {et~al.}(2014)\citenamefont
  {Chatrchyan} \emph {et~al.}}]{CMS:2013bza}%
  \BibitemOpen
  \bibfield  {author} {\bibinfo {author} {\bibfnamefont {S.}~\bibnamefont
  {Chatrchyan}} \emph {et~al.} (\bibinfo {collaboration} {CMS}),\ }\bibfield
  {title} {\bibinfo {title} {{Studies of azimuthal dihadron correlations in
  ultra-central PbPb collisions at $\sqrt{s_{NN}} =$ 2.76 TeV}},\ }\href
  {https://doi.org/10.1007/JHEP02(2014)088} {\bibfield  {journal} {\bibinfo
  {journal} {JHEP}\ }\textbf {\bibinfo {volume} {02}},\ \bibinfo {pages}
  {088}},\ \Eprint {https://arxiv.org/abs/1312.1845} {arXiv:1312.1845
  [nucl-ex]} \BibitemShut {NoStop}%
%%CITATION = ARXIV:1312.1845;%%
\bibitem [{\citenamefont {Khachatryan}\ \emph {et~al.}(2015)\citenamefont
  {Khachatryan} \emph {et~al.}}]{Khachatryan:2015oea}%
  \BibitemOpen
  \bibfield  {author} {\bibinfo {author} {\bibfnamefont {V.}~\bibnamefont
  {Khachatryan}} \emph {et~al.} (\bibinfo {collaboration} {CMS}),\ }\bibfield
  {title} {\bibinfo {title} {{Evidence for transverse momentum and
  pseudorapidity dependent event plane fluctuations in PbPb and pPb
  collisions}},\ }\href {https://doi.org/10.1103/PhysRevC.92.034911} {\bibfield
   {journal} {\bibinfo  {journal} {Phys. Rev.}\ }\textbf {\bibinfo {volume}
  {C92}},\ \bibinfo {pages} {034911} (\bibinfo {year} {2015})},\ \Eprint
  {https://arxiv.org/abs/1503.01692} {arXiv:1503.01692 [nucl-ex]} \BibitemShut
  {NoStop}%
%%CITATION = ARXIV:1503.01692;%%
\bibitem [{\citenamefont {Aamodt}\ \emph {et~al.}(2012)\citenamefont {Aamodt}
  \emph {et~al.}}]{Aamodt:2011by}%
  \BibitemOpen
  \bibfield  {author} {\bibinfo {author} {\bibfnamefont {K.}~\bibnamefont
  {Aamodt}} \emph {et~al.} (\bibinfo {collaboration} {ALICE}),\ }\bibfield
  {title} {\bibinfo {title} {{Harmonic decomposition of two-particle angular
  correlations in Pb-Pb collisions at $\sqrt{s_\mathrm{NN}}=$ 2.76 TeV}},\
  }\href {https://doi.org/10.1016/j.physletb.2012.01.060} {\bibfield  {journal}
  {\bibinfo  {journal} {Phys. Lett.}\ }\textbf {\bibinfo {volume} {B708}},\
  \bibinfo {pages} {249} (\bibinfo {year} {2012})},\ \Eprint
  {https://arxiv.org/abs/1109.2501} {arXiv:1109.2501 [nucl-ex]} \BibitemShut
  {NoStop}%
%%CITATION = ARXIV:1109.2501;%%
\bibitem [{\citenamefont {Acharya}\ \emph
  {et~al.}(2022{\natexlab{a}})\citenamefont {Acharya} \emph
  {et~al.}}]{ALICE:2022zks}%
  \BibitemOpen
  \bibfield  {author} {\bibinfo {author} {\bibfnamefont {S.}~\bibnamefont
  {Acharya}} \emph {et~al.} (\bibinfo {collaboration} {ALICE}),\ }\bibfield
  {title} {\bibinfo {title} {{Anisotropic flow and flow fluctuations of
  identified hadrons in Pb$-$Pb collisions at $\sqrt{s_{\mathrm{NN}}}$ = 5.02
  TeV}},\ }\href@noop {} {\  (\bibinfo {year} {2022}{\natexlab{a}})},\ \Eprint
  {https://arxiv.org/abs/2206.04587} {arXiv:2206.04587 [nucl-ex]} \BibitemShut
  {NoStop}%
\bibitem [{\citenamefont {Qiu}\ and\ \citenamefont {Heinz}(2011)}]{Qiu:2011iv}%
  \BibitemOpen
  \bibfield  {author} {\bibinfo {author} {\bibfnamefont {Z.}~\bibnamefont
  {Qiu}}\ and\ \bibinfo {author} {\bibfnamefont {U.~W.}\ \bibnamefont
  {Heinz}},\ }\bibfield  {title} {\bibinfo {title} {{Event-by-event shape and
  flow fluctuations of relativistic heavy-ion collision fireballs}},\ }\href
  {https://doi.org/10.1103/PhysRevC.84.024911} {\bibfield  {journal} {\bibinfo
  {journal} {Phys. Rev.}\ }\textbf {\bibinfo {volume} {C84}},\ \bibinfo {pages}
  {024911} (\bibinfo {year} {2011})},\ \Eprint
  {https://arxiv.org/abs/1104.0650} {arXiv:1104.0650 [nucl-th]} \BibitemShut
  {NoStop}%
%%CITATION = ARXIV:1104.0650;%%
\bibitem [{\citenamefont {Teaney}\ and\ \citenamefont
  {Yan}(2011)}]{Teaney:2010vd}%
  \BibitemOpen
  \bibfield  {author} {\bibinfo {author} {\bibfnamefont {D.}~\bibnamefont
  {Teaney}}\ and\ \bibinfo {author} {\bibfnamefont {L.}~\bibnamefont {Yan}},\
  }\bibfield  {title} {\bibinfo {title} {{Triangularity and Dipole Asymmetry in
  Heavy Ion Collisions}},\ }\href {https://doi.org/10.1103/PhysRevC.83.064904}
  {\bibfield  {journal} {\bibinfo  {journal} {Phys. Rev.}\ }\textbf {\bibinfo
  {volume} {C83}},\ \bibinfo {pages} {064904} (\bibinfo {year} {2011})},\
  \Eprint {https://arxiv.org/abs/1010.1876} {arXiv:1010.1876 [nucl-th]}
  \BibitemShut {NoStop}%
%%CITATION = ARXIV:1010.1876;%%
\bibitem [{\citenamefont {Jia}\ and\ \citenamefont
  {Mohapatra}(2013)}]{Jia:2012ma}%
  \BibitemOpen
  \bibfield  {author} {\bibinfo {author} {\bibfnamefont {J.}~\bibnamefont
  {Jia}}\ and\ \bibinfo {author} {\bibfnamefont {S.}~\bibnamefont
  {Mohapatra}},\ }\bibfield  {title} {\bibinfo {title} {{A Method for studying
  initial geometry fluctuations via event plane correlations in heavy ion
  collisions}},\ }\href {https://doi.org/10.1140/epjc/s10052-013-2510-y}
  {\bibfield  {journal} {\bibinfo  {journal} {Eur. Phys. J. C}\ }\textbf
  {\bibinfo {volume} {73}},\ \bibinfo {pages} {2510} (\bibinfo {year}
  {2013})},\ \Eprint {https://arxiv.org/abs/1203.5095} {arXiv:1203.5095
  [nucl-th]} \BibitemShut {NoStop}%
\bibitem [{\citenamefont {Aad}\ \emph {et~al.}(2014)\citenamefont {Aad} \emph
  {et~al.}}]{Aad:2014fla}%
  \BibitemOpen
  \bibfield  {author} {\bibinfo {author} {\bibfnamefont {G.}~\bibnamefont
  {Aad}} \emph {et~al.} (\bibinfo {collaboration} {ATLAS}),\ }\bibfield
  {title} {\bibinfo {title} {{Measurement of event-plane correlations in
  $\sqrt{s_{NN}}=2.76$ TeV lead-lead collisions with the ATLAS detector}},\
  }\href {https://doi.org/10.1103/PhysRevC.90.024905} {\bibfield  {journal}
  {\bibinfo  {journal} {Phys. Rev.}\ }\textbf {\bibinfo {volume} {C90}},\
  \bibinfo {pages} {024905} (\bibinfo {year} {2014})},\ \Eprint
  {https://arxiv.org/abs/1403.0489} {arXiv:1403.0489 [hep-ex]} \BibitemShut
  {NoStop}%
%%CITATION = ARXIV:1403.0489;%%
\bibitem [{\citenamefont {Qiu}\ and\ \citenamefont {Heinz}(2012)}]{Qiu:2012uy}%
  \BibitemOpen
  \bibfield  {author} {\bibinfo {author} {\bibfnamefont {Z.}~\bibnamefont
  {Qiu}}\ and\ \bibinfo {author} {\bibfnamefont {U.}~\bibnamefont {Heinz}},\
  }\bibfield  {title} {\bibinfo {title} {{Hydrodynamic event-plane correlations
  in Pb+Pb collisions at $\sqrt{s}=2.76$ATeV}},\ }\href
  {https://doi.org/10.1016/j.physletb.2012.09.030} {\bibfield  {journal}
  {\bibinfo  {journal} {Phys. Lett.}\ }\textbf {\bibinfo {volume} {B717}},\
  \bibinfo {pages} {261} (\bibinfo {year} {2012})},\ \Eprint
  {https://arxiv.org/abs/1208.1200} {arXiv:1208.1200 [nucl-th]} \BibitemShut
  {NoStop}%
%%CITATION = ARXIV:1208.1200;%%
\bibitem [{\citenamefont {Bo\.zek}(2018)}]{Bozek:2018nne}%
  \BibitemOpen
  \bibfield  {author} {\bibinfo {author} {\bibfnamefont {P.}~\bibnamefont
  {Bo\.zek}},\ }\bibfield  {title} {\bibinfo {title} {{Angle and magnitude
  decorrelation in the factorization breaking of collective flow}},\ }\href
  {https://doi.org/10.1103/PhysRevC.98.064906} {\bibfield  {journal} {\bibinfo
  {journal} {Phys. Rev. C}\ }\textbf {\bibinfo {volume} {98}},\ \bibinfo
  {pages} {064906} (\bibinfo {year} {2018})},\ \Eprint
  {https://arxiv.org/abs/1808.04248} {arXiv:1808.04248 [nucl-th]} \BibitemShut
  {NoStop}%
\bibitem [{\citenamefont {Gardim}\ \emph {et~al.}(2018)\citenamefont {Gardim},
  \citenamefont {Grassi}, \citenamefont {Ishida}, \citenamefont {Luzum},
  \citenamefont {Magalh\~aes},\ and\ \citenamefont
  {Noronha-Hostler}}]{Gardim:2017ruc}%
  \BibitemOpen
  \bibfield  {author} {\bibinfo {author} {\bibfnamefont {F.~G.}\ \bibnamefont
  {Gardim}}, \bibinfo {author} {\bibfnamefont {F.}~\bibnamefont {Grassi}},
  \bibinfo {author} {\bibfnamefont {P.}~\bibnamefont {Ishida}}, \bibinfo
  {author} {\bibfnamefont {M.}~\bibnamefont {Luzum}}, \bibinfo {author}
  {\bibfnamefont {P.~S.}\ \bibnamefont {Magalh\~aes}},\ and\ \bibinfo {author}
  {\bibfnamefont {J.}~\bibnamefont {Noronha-Hostler}},\ }\bibfield  {title}
  {\bibinfo {title} {{Sensitivity of observables to coarse-graining size in
  heavy-ion collisions}},\ }\href {https://doi.org/10.1103/PhysRevC.97.064919}
  {\bibfield  {journal} {\bibinfo  {journal} {Phys. Rev. C}\ }\textbf {\bibinfo
  {volume} {97}},\ \bibinfo {pages} {064919} (\bibinfo {year} {2018})},\
  \Eprint {https://arxiv.org/abs/1712.03912} {arXiv:1712.03912 [nucl-th]}
  \BibitemShut {NoStop}%
\bibitem [{\citenamefont {Zhao}\ \emph {et~al.}(2017)\citenamefont {Zhao},
  \citenamefont {Xu},\ and\ \citenamefont {Song}}]{Zhao:2017yhj}%
  \BibitemOpen
  \bibfield  {author} {\bibinfo {author} {\bibfnamefont {W.}~\bibnamefont
  {Zhao}}, \bibinfo {author} {\bibfnamefont {H.-j.}\ \bibnamefont {Xu}},\ and\
  \bibinfo {author} {\bibfnamefont {H.}~\bibnamefont {Song}},\ }\bibfield
  {title} {\bibinfo {title} {{Collective flow in 2.76 A TeV and 5.02 A TeV
  Pb+Pb collisions}},\ }\href {https://doi.org/10.1140/epjc/s10052-017-5186-x}
  {\bibfield  {journal} {\bibinfo  {journal} {Eur. Phys. J.}\ }\textbf
  {\bibinfo {volume} {C77}},\ \bibinfo {pages} {645} (\bibinfo {year}
  {2017})},\ \Eprint {https://arxiv.org/abs/1703.10792} {arXiv:1703.10792
  [nucl-th]} \BibitemShut {NoStop}%
%%CITATION = ARXIV:1703.10792;%%
\bibitem [{\citenamefont {Barbosa}\ \emph {et~al.}(2021)\citenamefont
  {Barbosa}, \citenamefont {Gardim}, \citenamefont {Grassi}, \citenamefont
  {Ishida}, \citenamefont {Luzum}, \citenamefont {Machado},\ and\ \citenamefont
  {Noronha-Hostler}}]{Barbosa:2021ccw}%
  \BibitemOpen
  \bibfield  {author} {\bibinfo {author} {\bibfnamefont {L.}~\bibnamefont
  {Barbosa}}, \bibinfo {author} {\bibfnamefont {F.~G.}\ \bibnamefont {Gardim}},
  \bibinfo {author} {\bibfnamefont {F.}~\bibnamefont {Grassi}}, \bibinfo
  {author} {\bibfnamefont {P.}~\bibnamefont {Ishida}}, \bibinfo {author}
  {\bibfnamefont {M.}~\bibnamefont {Luzum}}, \bibinfo {author} {\bibfnamefont
  {M.~V.}\ \bibnamefont {Machado}},\ and\ \bibinfo {author} {\bibfnamefont
  {J.}~\bibnamefont {Noronha-Hostler}},\ }\bibfield  {title} {\bibinfo {title}
  {{Predictions for flow harmonic distributions and flow factorization ratios
  at RHIC}},\ }\href@noop {} {\  (\bibinfo {year} {2021})},\ \Eprint
  {https://arxiv.org/abs/2105.12792} {arXiv:2105.12792 [nucl-th]} \BibitemShut
  {NoStop}%
\bibitem [{\citenamefont {Nielsen}(2021)}]{IS2021}%
  \BibitemOpen
  \bibfield  {author} {\bibinfo {author} {\bibfnamefont {E.~G.}\ \bibnamefont
  {Nielsen}} (\bibinfo {collaboration} {for the ALICE collaboration}),\
  }\bibfield  {title} {\bibinfo {title} {Fluctuations and correlations of flow
  in heavy-ion collisions measured by alice}} (\bibinfo {year} {2021}),\
  \bibinfo {note} {{T}he $\mathrm{VI^{th}}$ International Conference on the
  Initial Stages of High-Energy Nuclear Collisions,
  https://indico.cern.ch/event/854124/contributions/4134638/}\BibitemShut
  {NoStop}%
\bibitem [{\citenamefont {Bozek}\ and\ \citenamefont
  {Samanta}(2022)}]{Bozek:2021mov}%
  \BibitemOpen
  \bibfield  {author} {\bibinfo {author} {\bibfnamefont {P.}~\bibnamefont
  {Bozek}}\ and\ \bibinfo {author} {\bibfnamefont {R.}~\bibnamefont
  {Samanta}},\ }\bibfield  {title} {\bibinfo {title} {{Factorization breaking
  for higher moments of harmonic flow}},\ }\href
  {https://doi.org/10.1103/PhysRevC.105.034904} {\bibfield  {journal} {\bibinfo
   {journal} {Phys. Rev. C}\ }\textbf {\bibinfo {volume} {105}},\ \bibinfo
  {pages} {034904} (\bibinfo {year} {2022})},\ \Eprint
  {https://arxiv.org/abs/2109.07781} {arXiv:2109.07781 [nucl-th]} \BibitemShut
  {NoStop}%
\bibitem [{\citenamefont {Acharya}\ \emph
  {et~al.}(2022{\natexlab{b}})\citenamefont {Acharya} \emph
  {et~al.}}]{ALICE:2022smy}%
  \BibitemOpen
  \bibfield  {author} {\bibinfo {author} {\bibfnamefont {S.}~\bibnamefont
  {Acharya}} \emph {et~al.} (\bibinfo {collaboration} {ALICE}),\ }\bibfield
  {title} {\bibinfo {title} {{Observation of flow angle and flow magnitude
  fluctuations in Pb-Pb collisions at $\sqrt{s_{\rm NN}}$ = 5.02 TeV at the
  LHC}},\ }\href@noop {} {\  (\bibinfo {year} {2022}{\natexlab{b}})},\ \Eprint
  {https://arxiv.org/abs/2206.04574} {arXiv:2206.04574 [nucl-ex]} \BibitemShut
  {NoStop}%
\bibitem [{\citenamefont {Lin}\ \emph {et~al.}(2005)\citenamefont {Lin},
  \citenamefont {Ko}, \citenamefont {Li}, \citenamefont {Zhang},\ and\
  \citenamefont {Pal}}]{Lin:2004en}%
  \BibitemOpen
  \bibfield  {author} {\bibinfo {author} {\bibfnamefont {Z.-W.}\ \bibnamefont
  {Lin}}, \bibinfo {author} {\bibfnamefont {C.~M.}\ \bibnamefont {Ko}},
  \bibinfo {author} {\bibfnamefont {B.-A.}\ \bibnamefont {Li}}, \bibinfo
  {author} {\bibfnamefont {B.}~\bibnamefont {Zhang}},\ and\ \bibinfo {author}
  {\bibfnamefont {S.}~\bibnamefont {Pal}},\ }\bibfield  {title} {\bibinfo
  {title} {{A Multi-phase transport model for relativistic heavy ion
  collisions}},\ }\href {https://doi.org/10.1103/PhysRevC.72.064901} {\bibfield
   {journal} {\bibinfo  {journal} {Phys. Rev.}\ }\textbf {\bibinfo {volume}
  {C72}},\ \bibinfo {pages} {064901} (\bibinfo {year} {2005})},\ \Eprint
  {https://arxiv.org/abs/nucl-th/0411110} {arXiv:nucl-th/0411110 [nucl-th]}
  \BibitemShut {NoStop}%
%%CITATION = NUCL-TH/0411110;%%
\bibitem [{\citenamefont {Lin}\ and\ \citenamefont
  {Zheng}(2021)}]{Lin:2021mdn}%
  \BibitemOpen
  \bibfield  {author} {\bibinfo {author} {\bibfnamefont {Z.-W.}\ \bibnamefont
  {Lin}}\ and\ \bibinfo {author} {\bibfnamefont {L.}~\bibnamefont {Zheng}},\
  }\bibfield  {title} {\bibinfo {title} {{Further developments of a multi-phase
  transport model for relativistic nuclear collisions}},\ }\href
  {https://doi.org/10.1007/s41365-021-00944-5} {\bibfield  {journal} {\bibinfo
  {journal} {Nucl. Sci. Tech.}\ }\textbf {\bibinfo {volume} {32}},\ \bibinfo
  {pages} {113} (\bibinfo {year} {2021})},\ \Eprint
  {https://arxiv.org/abs/2110.02989} {arXiv:2110.02989 [nucl-th]} \BibitemShut
  {NoStop}%
\bibitem [{\citenamefont {Lin}\ and\ \citenamefont {Ko}(2002)}]{Lin:2001zk}%
  \BibitemOpen
  \bibfield  {author} {\bibinfo {author} {\bibfnamefont {Z.-w.}\ \bibnamefont
  {Lin}}\ and\ \bibinfo {author} {\bibfnamefont {C.~M.}\ \bibnamefont {Ko}},\
  }\bibfield  {title} {\bibinfo {title} {{Partonic effects on the elliptic flow
  at RHIC}},\ }\href {https://doi.org/10.1103/PhysRevC.65.034904} {\bibfield
  {journal} {\bibinfo  {journal} {Phys. Rev. C}\ }\textbf {\bibinfo {volume}
  {65}},\ \bibinfo {pages} {034904} (\bibinfo {year} {2002})},\ \Eprint
  {https://arxiv.org/abs/nucl-th/0108039} {arXiv:nucl-th/0108039} \BibitemShut
  {NoStop}%
\bibitem [{\citenamefont {Xu}\ and\ \citenamefont
  {Ko}(2011{\natexlab{a}})}]{Xu:2011fe}%
  \BibitemOpen
  \bibfield  {author} {\bibinfo {author} {\bibfnamefont {J.}~\bibnamefont
  {Xu}}\ and\ \bibinfo {author} {\bibfnamefont {C.~M.}\ \bibnamefont {Ko}},\
  }\bibfield  {title} {\bibinfo {title} {{Triangular flow in heavy ion
  collisions in a multiphase transport model}},\ }\href
  {https://doi.org/10.1103/PhysRevC.84.014903} {\bibfield  {journal} {\bibinfo
  {journal} {Phys. Rev. C}\ }\textbf {\bibinfo {volume} {84}},\ \bibinfo
  {pages} {014903} (\bibinfo {year} {2011}{\natexlab{a}})},\ \Eprint
  {https://arxiv.org/abs/1103.5187} {arXiv:1103.5187 [nucl-th]} \BibitemShut
  {NoStop}%
\bibitem [{\citenamefont {Xu}\ and\ \citenamefont
  {Ko}(2011{\natexlab{b}})}]{Xu:2011fi}%
  \BibitemOpen
  \bibfield  {author} {\bibinfo {author} {\bibfnamefont {J.}~\bibnamefont
  {Xu}}\ and\ \bibinfo {author} {\bibfnamefont {C.~M.}\ \bibnamefont {Ko}},\
  }\bibfield  {title} {\bibinfo {title} {{Pb-Pb collisions at
  $\sqrt{s_{NN}}=2.76$ TeV in a multiphase transport model}},\ }\href
  {https://doi.org/10.1103/PhysRevC.83.034904} {\bibfield  {journal} {\bibinfo
  {journal} {Phys. Rev. C}\ }\textbf {\bibinfo {volume} {83}},\ \bibinfo
  {pages} {034904} (\bibinfo {year} {2011}{\natexlab{b}})},\ \Eprint
  {https://arxiv.org/abs/1101.2231} {arXiv:1101.2231 [nucl-th]} \BibitemShut
  {NoStop}%
\bibitem [{\citenamefont {Feng}\ \emph {et~al.}(2017)\citenamefont {Feng},
  \citenamefont {Huang},\ and\ \citenamefont {Liu}}]{Feng:2016emh}%
  \BibitemOpen
  \bibfield  {author} {\bibinfo {author} {\bibfnamefont {Z.}~\bibnamefont
  {Feng}}, \bibinfo {author} {\bibfnamefont {G.-M.}\ \bibnamefont {Huang}},\
  and\ \bibinfo {author} {\bibfnamefont {F.}~\bibnamefont {Liu}},\ }\bibfield
  {title} {\bibinfo {title} {{Anisotropic flow of Pb+Pb $\sqrt{s_{\rm NN}}$ =
  5.02 TeV from a Multi-Phase Transport Model}},\ }\href
  {https://doi.org/10.1088/1674-1137/41/2/024001} {\bibfield  {journal}
  {\bibinfo  {journal} {Chin. Phys. C}\ }\textbf {\bibinfo {volume} {41}},\
  \bibinfo {pages} {024001} (\bibinfo {year} {2017})},\ \Eprint
  {https://arxiv.org/abs/1606.02416} {arXiv:1606.02416 [nucl-ex]} \BibitemShut
  {NoStop}%
\bibitem [{\citenamefont {Wang}\ and\ \citenamefont
  {Gyulassy}(1991{\natexlab{a}})}]{PhysRevD.44.3501}%
  \BibitemOpen
  \bibfield  {author} {\bibinfo {author} {\bibfnamefont {X.-N.}\ \bibnamefont
  {Wang}}\ and\ \bibinfo {author} {\bibfnamefont {M.}~\bibnamefont
  {Gyulassy}},\ }\bibfield  {title} {\bibinfo {title} {hijing: A monte carlo
  model for multiple jet production in $\mathrm{pp}$, $\mathrm{pA}$, and
  $\mathrm{AA}$ collisions},\ }\href {https://doi.org/10.1103/PhysRevD.44.3501}
  {\bibfield  {journal} {\bibinfo  {journal} {Phys. Rev.}\ }\textbf {\bibinfo
  {volume} {D44}},\ \bibinfo {pages} {3501} (\bibinfo {year}
  {1991}{\natexlab{a}})}\BibitemShut {NoStop}%
\bibitem [{\citenamefont {Zhang}(1998)}]{Zhang:1997ej}%
  \BibitemOpen
  \bibfield  {author} {\bibinfo {author} {\bibfnamefont {B.}~\bibnamefont
  {Zhang}},\ }\bibfield  {title} {\bibinfo {title} {{ZPC 1.0.1: A Parton
  cascade for ultrarelativistic heavy ion collisions}},\ }\href
  {https://doi.org/10.1016/S0010-4655(98)00010-1} {\bibfield  {journal}
  {\bibinfo  {journal} {Comput. Phys. Commun.}\ }\textbf {\bibinfo {volume}
  {109}},\ \bibinfo {pages} {193} (\bibinfo {year} {1998})},\ \Eprint
  {https://arxiv.org/abs/nucl-th/9709009} {arXiv:nucl-th/9709009} \BibitemShut
  {NoStop}%
\bibitem [{\citenamefont {Lin}(2014)}]{Lin:2014tya}%
  \BibitemOpen
  \bibfield  {author} {\bibinfo {author} {\bibfnamefont {Z.-W.}\ \bibnamefont
  {Lin}},\ }\bibfield  {title} {\bibinfo {title} {{Evolution of transverse flow
  and effective temperatures in the parton phase from a multi-phase transport
  model}},\ }\href {https://doi.org/10.1103/PhysRevC.90.014904} {\bibfield
  {journal} {\bibinfo  {journal} {Phys. Rev. C}\ }\textbf {\bibinfo {volume}
  {90}},\ \bibinfo {pages} {014904} (\bibinfo {year} {2014})},\ \Eprint
  {https://arxiv.org/abs/1403.6321} {arXiv:1403.6321 [nucl-th]} \BibitemShut
  {NoStop}%
\bibitem [{\citenamefont {Chen}\ and\ \citenamefont {Ko}(2006)}]{Chen:2005mr}%
  \BibitemOpen
  \bibfield  {author} {\bibinfo {author} {\bibfnamefont {L.-W.}\ \bibnamefont
  {Chen}}\ and\ \bibinfo {author} {\bibfnamefont {C.~M.}\ \bibnamefont {Ko}},\
  }\bibfield  {title} {\bibinfo {title} {{System size dependence of elliptic
  flows in relativistic heavy-ion collisions}},\ }\href
  {https://doi.org/10.1016/j.physletb.2006.01.037} {\bibfield  {journal}
  {\bibinfo  {journal} {Phys. Lett.}\ }\textbf {\bibinfo {volume} {B634}},\
  \bibinfo {pages} {205} (\bibinfo {year} {2006})},\ \Eprint
  {https://arxiv.org/abs/nucl-th/0505044} {arXiv:nucl-th/0505044 [nucl-th]}
  \BibitemShut {NoStop}%
%%CITATION = NUCL-TH/0505044;%%
\bibitem [{\citenamefont {Li}\ and\ \citenamefont {Ko}(1995)}]{Li:1995pra}%
  \BibitemOpen
  \bibfield  {author} {\bibinfo {author} {\bibfnamefont {B.-A.}\ \bibnamefont
  {Li}}\ and\ \bibinfo {author} {\bibfnamefont {C.~M.}\ \bibnamefont {Ko}},\
  }\bibfield  {title} {\bibinfo {title} {{Formation of superdense hadronic
  matter in high-energy heavy ion collisions}},\ }\href
  {https://doi.org/10.1103/PhysRevC.52.2037} {\bibfield  {journal} {\bibinfo
  {journal} {Phys. Rev.}\ }\textbf {\bibinfo {volume} {C52}},\ \bibinfo {pages}
  {2037} (\bibinfo {year} {1995})},\ \Eprint
  {https://arxiv.org/abs/nucl-th/9505016} {arXiv:nucl-th/9505016 [nucl-th]}
  \BibitemShut {NoStop}%
%%CITATION = NUCL-TH/9505016;%%
\bibitem [{\citenamefont {Li}\ \emph {et~al.}(2001)\citenamefont {Li},
  \citenamefont {Sustich}, \citenamefont {Zhang},\ and\ \citenamefont
  {Ko}}]{Li:2001xh}%
  \BibitemOpen
  \bibfield  {author} {\bibinfo {author} {\bibfnamefont {B.}~\bibnamefont
  {Li}}, \bibinfo {author} {\bibfnamefont {A.~T.}\ \bibnamefont {Sustich}},
  \bibinfo {author} {\bibfnamefont {B.}~\bibnamefont {Zhang}},\ and\ \bibinfo
  {author} {\bibfnamefont {C.~M.}\ \bibnamefont {Ko}},\ }\bibfield  {title}
  {\bibinfo {title} {{Studies of superdense hadronic matter in a relativistic
  transport model}},\ }\href {https://doi.org/10.1142/S0218301301000575}
  {\bibfield  {journal} {\bibinfo  {journal} {Int. J. Mod. Phys. E}\ }\textbf
  {\bibinfo {volume} {10}},\ \bibinfo {pages} {267} (\bibinfo {year}
  {2001})}\BibitemShut {NoStop}%
\bibitem [{\citenamefont {Wang}\ and\ \citenamefont
  {Gyulassy}(1991{\natexlab{b}})}]{Wang:1991hta}%
  \BibitemOpen
  \bibfield  {author} {\bibinfo {author} {\bibfnamefont {X.-N.}\ \bibnamefont
  {Wang}}\ and\ \bibinfo {author} {\bibfnamefont {M.}~\bibnamefont
  {Gyulassy}},\ }\bibfield  {title} {\bibinfo {title} {{HIJING: A Monte Carlo
  model for multiple jet production in p p, p A and A A collisions}},\ }\href
  {https://doi.org/10.1103/PhysRevD.44.3501} {\bibfield  {journal} {\bibinfo
  {journal} {Phys.\ Rev.\ D}\ }\textbf {\bibinfo {volume} {44}},\ \bibinfo
  {pages} {3501} (\bibinfo {year} {1991}{\natexlab{b}})}\BibitemShut {NoStop}%
\bibitem [{\citenamefont {Gyulassy}\ and\ \citenamefont
  {Wang}(1994)}]{Gyulassy:1994ew}%
  \BibitemOpen
  \bibfield  {author} {\bibinfo {author} {\bibfnamefont {M.}~\bibnamefont
  {Gyulassy}}\ and\ \bibinfo {author} {\bibfnamefont {X.-N.}\ \bibnamefont
  {Wang}},\ }\bibfield  {title} {\bibinfo {title} {{HIJING 1.0: A Monte Carlo
  program for parton and particle production in high-energy hadronic and
  nuclear collisions}},\ }\href {https://doi.org/10.1016/0010-4655(94)90057-4}
  {\bibfield  {journal} {\bibinfo  {journal} {Comput. Phys. Commun.}\ }\textbf
  {\bibinfo {volume} {83}},\ \bibinfo {pages} {307} (\bibinfo {year} {1994})},\
  \Eprint {https://arxiv.org/abs/nucl-th/9502021} {arXiv:nucl-th/9502021
  [nucl-th]} \BibitemShut {NoStop}%
%%CITATION = NUCL-TH/9502021;%%
\bibitem [{\citenamefont {Gyulassy}\ \emph {et~al.}(1997)\citenamefont
  {Gyulassy}, \citenamefont {Pang},\ and\ \citenamefont
  {Zhang}}]{Gyulassy:1997ib}%
  \BibitemOpen
  \bibfield  {author} {\bibinfo {author} {\bibfnamefont {M.}~\bibnamefont
  {Gyulassy}}, \bibinfo {author} {\bibfnamefont {Y.}~\bibnamefont {Pang}},\
  and\ \bibinfo {author} {\bibfnamefont {B.}~\bibnamefont {Zhang}},\ }\bibfield
   {title} {\bibinfo {title} {{Transverse energy evolution as a test of parton
  cascade models}},\ }\href {https://doi.org/10.1016/S0375-9474(97)00604-0}
  {\bibfield  {journal} {\bibinfo  {journal} {Nucl. Phys. A}\ }\textbf
  {\bibinfo {volume} {626}},\ \bibinfo {pages} {999} (\bibinfo {year}
  {1997})},\ \Eprint {https://arxiv.org/abs/nucl-th/9709025}
  {arXiv:nucl-th/9709025} \BibitemShut {NoStop}%
\bibitem [{\citenamefont {Zhang}\ \emph {et~al.}(1999)\citenamefont {Zhang},
  \citenamefont {Gyulassy},\ and\ \citenamefont {Ko}}]{Zhang:1999rs}%
  \BibitemOpen
  \bibfield  {author} {\bibinfo {author} {\bibfnamefont {B.}~\bibnamefont
  {Zhang}}, \bibinfo {author} {\bibfnamefont {M.}~\bibnamefont {Gyulassy}},\
  and\ \bibinfo {author} {\bibfnamefont {C.~M.}\ \bibnamefont {Ko}},\
  }\bibfield  {title} {\bibinfo {title} {{Elliptic flow from a parton
  cascade}},\ }\href {https://doi.org/10.1016/S0370-2693(99)00456-6} {\bibfield
   {journal} {\bibinfo  {journal} {Phys. Lett. B}\ }\textbf {\bibinfo {volume}
  {455}},\ \bibinfo {pages} {45} (\bibinfo {year} {1999})},\ \Eprint
  {https://arxiv.org/abs/nucl-th/9902016} {arXiv:nucl-th/9902016} \BibitemShut
  {NoStop}%
\bibitem [{\citenamefont {Borghini}\ \emph {et~al.}(2001)\citenamefont
  {Borghini}, \citenamefont {Dinh},\ and\ \citenamefont
  {Ollitrault}}]{Borghini:2001vi}%
  \BibitemOpen
  \bibfield  {author} {\bibinfo {author} {\bibfnamefont {N.}~\bibnamefont
  {Borghini}}, \bibinfo {author} {\bibfnamefont {P.~M.}\ \bibnamefont {Dinh}},\
  and\ \bibinfo {author} {\bibfnamefont {J.-Y.}\ \bibnamefont {Ollitrault}},\
  }\bibfield  {title} {\bibinfo {title} {{Flow analysis from multiparticle
  azimuthal correlations}},\ }\href
  {https://doi.org/10.1103/PhysRevC.64.054901} {\bibfield  {journal} {\bibinfo
  {journal} {Phys. Rev. C}\ }\textbf {\bibinfo {volume} {64}},\ \bibinfo
  {pages} {054901} (\bibinfo {year} {2001})},\ \Eprint
  {https://arxiv.org/abs/nucl-th/0105040} {arXiv:nucl-th/0105040} \BibitemShut
  {NoStop}%
\bibitem [{\citenamefont {Bilandzic}\ \emph {et~al.}(2011)\citenamefont
  {Bilandzic}, \citenamefont {Snellings},\ and\ \citenamefont
  {Voloshin}}]{Bilandzic:2010jr}%
  \BibitemOpen
  \bibfield  {author} {\bibinfo {author} {\bibfnamefont {A.}~\bibnamefont
  {Bilandzic}}, \bibinfo {author} {\bibfnamefont {R.}~\bibnamefont
  {Snellings}},\ and\ \bibinfo {author} {\bibfnamefont {S.}~\bibnamefont
  {Voloshin}},\ }\bibfield  {title} {\bibinfo {title} {{Flow analysis with
  cumulants: Direct calculations}},\ }\href
  {https://doi.org/10.1103/PhysRevC.83.044913} {\bibfield  {journal} {\bibinfo
  {journal} {Phys. Rev.}\ }\textbf {\bibinfo {volume} {C83}},\ \bibinfo {pages}
  {044913} (\bibinfo {year} {2011})},\ \Eprint
  {https://arxiv.org/abs/1010.0233} {arXiv:1010.0233 [nucl-ex]} \BibitemShut
  {NoStop}%
%%CITATION = ARXIV:1010.0233;%%
\bibitem [{\citenamefont {Agakishiev}\ \emph {et~al.}(2012)\citenamefont
  {Agakishiev} \emph {et~al.}}]{STAR:2011ert}%
  \BibitemOpen
  \bibfield  {author} {\bibinfo {author} {\bibfnamefont {G.}~\bibnamefont
  {Agakishiev}} \emph {et~al.} (\bibinfo {collaboration} {STAR}),\ }\bibfield
  {title} {\bibinfo {title} {{Energy and system-size dependence of two- and
  four-particle $v_2$ measurements in heavy-ion collisions at RHIC and their
  implications on flow fluctuations and nonflow}},\ }\href
  {https://doi.org/10.1103/PhysRevC.86.014904} {\bibfield  {journal} {\bibinfo
  {journal} {Phys. Rev. C}\ }\textbf {\bibinfo {volume} {86}},\ \bibinfo
  {pages} {014904} (\bibinfo {year} {2012})},\ \Eprint
  {https://arxiv.org/abs/1111.5637} {arXiv:1111.5637 [nucl-ex]} \BibitemShut
  {NoStop}%
\bibitem [{\citenamefont {Bilandzic}\ \emph {et~al.}(2014)\citenamefont
  {Bilandzic}, \citenamefont {Christensen}, \citenamefont {Gulbrandsen},
  \citenamefont {Hansen},\ and\ \citenamefont {Zhou}}]{Bilandzic:2013kga}%
  \BibitemOpen
  \bibfield  {author} {\bibinfo {author} {\bibfnamefont {A.}~\bibnamefont
  {Bilandzic}}, \bibinfo {author} {\bibfnamefont {C.~H.}\ \bibnamefont
  {Christensen}}, \bibinfo {author} {\bibfnamefont {K.}~\bibnamefont
  {Gulbrandsen}}, \bibinfo {author} {\bibfnamefont {A.}~\bibnamefont
  {Hansen}},\ and\ \bibinfo {author} {\bibfnamefont {Y.}~\bibnamefont {Zhou}},\
  }\bibfield  {title} {\bibinfo {title} {{Generic framework for anisotropic
  flow analyses with multiparticle azimuthal correlations}},\ }\href
  {https://doi.org/10.1103/PhysRevC.89.064904} {\bibfield  {journal} {\bibinfo
  {journal} {Phys. Rev.}\ }\textbf {\bibinfo {volume} {C89}},\ \bibinfo {pages}
  {064904} (\bibinfo {year} {2014})},\ \Eprint
  {https://arxiv.org/abs/1312.3572} {arXiv:1312.3572 [nucl-ex]} \BibitemShut
  {NoStop}%
%%CITATION = ARXIV:1312.3572;%%
\bibitem [{\citenamefont {Huo}\ \emph {et~al.}(2018)\citenamefont {Huo},
  \citenamefont {Gajdosov}, \citenamefont {Jia},\ and\ \citenamefont
  {Zhou}}]{Huo:2017nms}%
  \BibitemOpen
  \bibfield  {author} {\bibinfo {author} {\bibfnamefont {P.}~\bibnamefont
  {Huo}}, \bibinfo {author} {\bibfnamefont {K.}~\bibnamefont {Gajdosov}},
  \bibinfo {author} {\bibfnamefont {J.}~\bibnamefont {Jia}},\ and\ \bibinfo
  {author} {\bibfnamefont {Y.}~\bibnamefont {Zhou}},\ }\bibfield  {title}
  {\bibinfo {title} {{Importance of non-flow in mixed-harmonic multi-particle
  correlations in small collision systems}},\ }\href
  {https://doi.org/10.1016/j.physletb.2017.12.035} {\bibfield  {journal}
  {\bibinfo  {journal} {Phys. Lett.}\ }\textbf {\bibinfo {volume} {B777}},\
  \bibinfo {pages} {201} (\bibinfo {year} {2018})},\ \Eprint
  {https://arxiv.org/abs/1710.07567} {arXiv:1710.07567 [nucl-ex]} \BibitemShut
  {NoStop}%
%%CITATION = ARXIV:1710.07567;%%
\bibitem [{\citenamefont {Magdy}(2022)}]{Magdy:2022jai}%
  \BibitemOpen
  \bibfield  {author} {\bibinfo {author} {\bibfnamefont {N.}~\bibnamefont
  {Magdy}},\ }\bibfield  {title} {\bibinfo {title} {{Measuring differential
  flow angle fluctuations in relativistic nuclear collisions}},\ }\href
  {https://doi.org/10.1103/PhysRevC.106.044911} {\bibfield  {journal} {\bibinfo
   {journal} {Phys. Rev. C}\ }\textbf {\bibinfo {volume} {106}},\ \bibinfo
  {pages} {044911} (\bibinfo {year} {2022})},\ \Eprint
  {https://arxiv.org/abs/2207.04530} {arXiv:2207.04530 [nucl-th]} \BibitemShut
  {NoStop}%
\bibitem [{\citenamefont {Acharya}\ \emph {et~al.}(2019)\citenamefont {Acharya}
  \emph {et~al.}}]{ALICE:2019zfl}%
  \BibitemOpen
  \bibfield  {author} {\bibinfo {author} {\bibfnamefont {S.}~\bibnamefont
  {Acharya}} \emph {et~al.} (\bibinfo {collaboration} {ALICE}),\ }\bibfield
  {title} {\bibinfo {title} {{Investigations of Anisotropic Flow Using
  Multiparticle Azimuthal Correlations in pp, p-Pb, Xe-Xe, and Pb-Pb Collisions
  at the LHC}},\ }\href {https://doi.org/10.1103/PhysRevLett.123.142301}
  {\bibfield  {journal} {\bibinfo  {journal} {Phys. Rev. Lett.}\ }\textbf
  {\bibinfo {volume} {123}},\ \bibinfo {pages} {142301} (\bibinfo {year}
  {2019})},\ \Eprint {https://arxiv.org/abs/1903.01790} {arXiv:1903.01790
  [nucl-ex]} \BibitemShut {NoStop}%
\bibitem [{\citenamefont {Schenke}\ \emph {et~al.}(2011)\citenamefont
  {Schenke}, \citenamefont {Jeon},\ and\ \citenamefont
  {Gale}}]{Schenke:2010rr}%
  \BibitemOpen
  \bibfield  {author} {\bibinfo {author} {\bibfnamefont {B.}~\bibnamefont
  {Schenke}}, \bibinfo {author} {\bibfnamefont {S.}~\bibnamefont {Jeon}},\ and\
  \bibinfo {author} {\bibfnamefont {C.}~\bibnamefont {Gale}},\ }\bibfield
  {title} {\bibinfo {title} {{Elliptic and triangular flow in event-by-event
  (3+1)D viscous hydrodynamics}},\ }\href
  {https://doi.org/10.1103/PhysRevLett.106.042301} {\bibfield  {journal}
  {\bibinfo  {journal} {Phys. Rev. Lett.}\ }\textbf {\bibinfo {volume} {106}},\
  \bibinfo {pages} {042301} (\bibinfo {year} {2011})},\ \Eprint
  {https://arxiv.org/abs/1009.3244} {arXiv:1009.3244 [hep-ph]} \BibitemShut
  {NoStop}%
%%CITATION = ARXIV:1009.3244;%%
\bibitem [{\citenamefont {Acharya}\ \emph
  {et~al.}(2017{\natexlab{c}})\citenamefont {Acharya} \emph
  {et~al.}}]{Acharya:2017ino}%
  \BibitemOpen
  \bibfield  {author} {\bibinfo {author} {\bibfnamefont {S.}~\bibnamefont
  {Acharya}} \emph {et~al.} (\bibinfo {collaboration} {ALICE}),\ }\bibfield
  {title} {\bibinfo {title} {{Searches for transverse momentum dependent flow
  vector fluctuations in Pb-Pb and p-Pb collisions at the LHC}},\ }\href
  {https://doi.org/10.1007/JHEP09(2017)032} {\bibfield  {journal} {\bibinfo
  {journal} {JHEP}\ }\textbf {\bibinfo {volume} {09}},\ \bibinfo {pages}
  {032}},\ \Eprint {https://arxiv.org/abs/1707.05690} {arXiv:1707.05690
  [nucl-ex]} \BibitemShut {NoStop}%
%%CITATION = ARXIV:1707.05690;%%
\end{thebibliography}%

\end{document}